\documentclass{article} % For LaTeX2e
\usepackage{iclr2026_conference,times}

\usepackage{amsmath,amsfonts,bm}

\def\eqref#1{equation~\ref{#1}}
\def\1{\bm{1}}

\def\vg{{\bm{g}}}

\def\vr{{\bm{r}}}

\def\vw{{\bm{w}}}

\DeclareMathAlphabet{\mathsfit}{\encodingdefault}{\sfdefault}{m}{sl}
\SetMathAlphabet{\mathsfit}{bold}{\encodingdefault}{\sfdefault}{bx}{n}

\usepackage{amsmath,amsfonts,amssymb}
\usepackage{hyperref}
\usepackage{url}
\usepackage{booktabs}
\usepackage{multirow}
\usepackage{graphicx}
\usepackage{subcaption}
\usepackage{lipsum}
\usepackage{wrapfig}
\usepackage{amsthm}
\newtheorem{definition}{Definition}[section]
\usepackage{algorithm}
\usepackage{algpseudocode}
\usepackage{makecell}
\usepackage{enumitem}

\usepackage[draft,inline,nomargin,index]{fixme}
\fxsetup{theme=color,mode=multiuser}

\usepackage{shortcuts}

\iclrfinalcopy

\title{Defending MoE LLMs against Harmful Fine-Tuning via Safety Routing Alignment}

\author{Jaehan Kim, Minkyoo Song, Seungwon Shin, Sooel Son\thanks{Corresponding author}\\
KAIST\\
\texttt{\{jaehan,minkyoo9,claude,sl.son\}@kaist.ac.kr} \\}

\begin{document}

\maketitle

\begin{abstract}
Recent large language models (LLMs) have increasingly adopted the Mixture-of-Experts (MoE) architecture for efficiency. MoE-based LLMs heavily depend on a superficial safety mechanism in which harmful inputs are routed safety-critical experts. However, our analysis reveals that routing decisions for harmful inputs drift significantly after fine-tuning, exposing a critical vulnerability to harmful fine-tuning (HFT) attacks. Existing defenses, primarily designed for monolithic LLMs, are less effective for MoE LLMs as they fail to prevent drift in harmful input routing. To address this limitation, we propose \ours{}, a safe fine-tuning method tailored to MoE LLMs. \ours{} directly mitigates routing drift by penalizing the gap between the routing weights of a fine-tuned model and those of the initial safety-aligned model, thereby preserving the safety-aligned routing of harmful inputs to safety-critical experts. Experiments on open-source MoE LLMs ranging from 7B to 141B parameters demonstrate that \ours{} effectively mitigates HFT attacks, reducing the harmfulness score of OLMoE from 62.0 to 5.0, for example, while maintaining task utility within 1\% degradation and incurring only 2\% overhead. It significantly outperforms state-of-the-art defense methods for safeguarding LLM fine-tuning and remains effective in recent large-scale MoE LLMs such as gpt-oss and Llama 4. Our implementation is available at \url{https://anonymous.4open.science/r/SafeMoE}.

\end{abstract}

\section{Introduction}

Mixture-of-Experts (MoE)~\citep{shazeer2017outrageously} is a sparse model architecture that improves efficiency by dynamically routing inputs to a subset of expert layers, which have gained adoption for large language models (LLMs). Recent MoE-based LLMs, including gpt-oss~\citep{openai2025gptoss}, Llama 4~\citep{meta2025llama4}, Qwen3 MoE~\citep{qwen3technicalreport}, and DeepSeek-R1~\citep{guo2025deepseek}, have achieved surpassing performance on a wide range of challenging tasks, outperforming their monolithic counterparts.
However, recent studies~\citep{lai2025safex,fayyaz2025steering} show that the safety of MoE LLMs heavily relies on certain \textit{safety-critical} experts and intentionally manipulating routing decisions to disable these experts leads to significant increases in harmfulness. This superficial safety mechanism leaves MoE LLMs particularly susceptible to harmful fine-tuning (HFT) attacks~\citep{qi2024fine,yang2024shadow,zhan2024removing}. These attacks are designed to compromise the safety of a target LLM by injecting only a limited number of harmful samples into the training dataset, rendering a practical yet severe threat to commercial LLM providers given the growing prevalence of fine-tuning API services~\citep{openaifinetuning,geminifinetuning}.

Our systematic analysis of MoE LLMs uncovers a novel architectural vulnerability in their routing mechanism, which determines which expert layers should be activated for processing inputs.
We find that routing decisions for harmful inputs drift substantially from those of the initial safety-aligned model under both harmful and benign fine-tuning, a phenomenon we term \textit{safety routing drift}.
This drift impedes the activation of safety-critical experts and thereby undermines the model’s safety.
Given the reliance of MoE LLM safety on routing aligned toward these safety-critical experts, preserving the initial routing decisions of the safety-aligned models is important to safeguarding MoE LLMs against HFT attacks.
\begin{wrapfigure}{r}{0.654\linewidth}
% \vspace{-0.2cm}
% \begin{figure}[ht]
    \centering
     \begin{subfigure}[t]{0.495\linewidth}
        \centering
        \includegraphics[width=\linewidth]{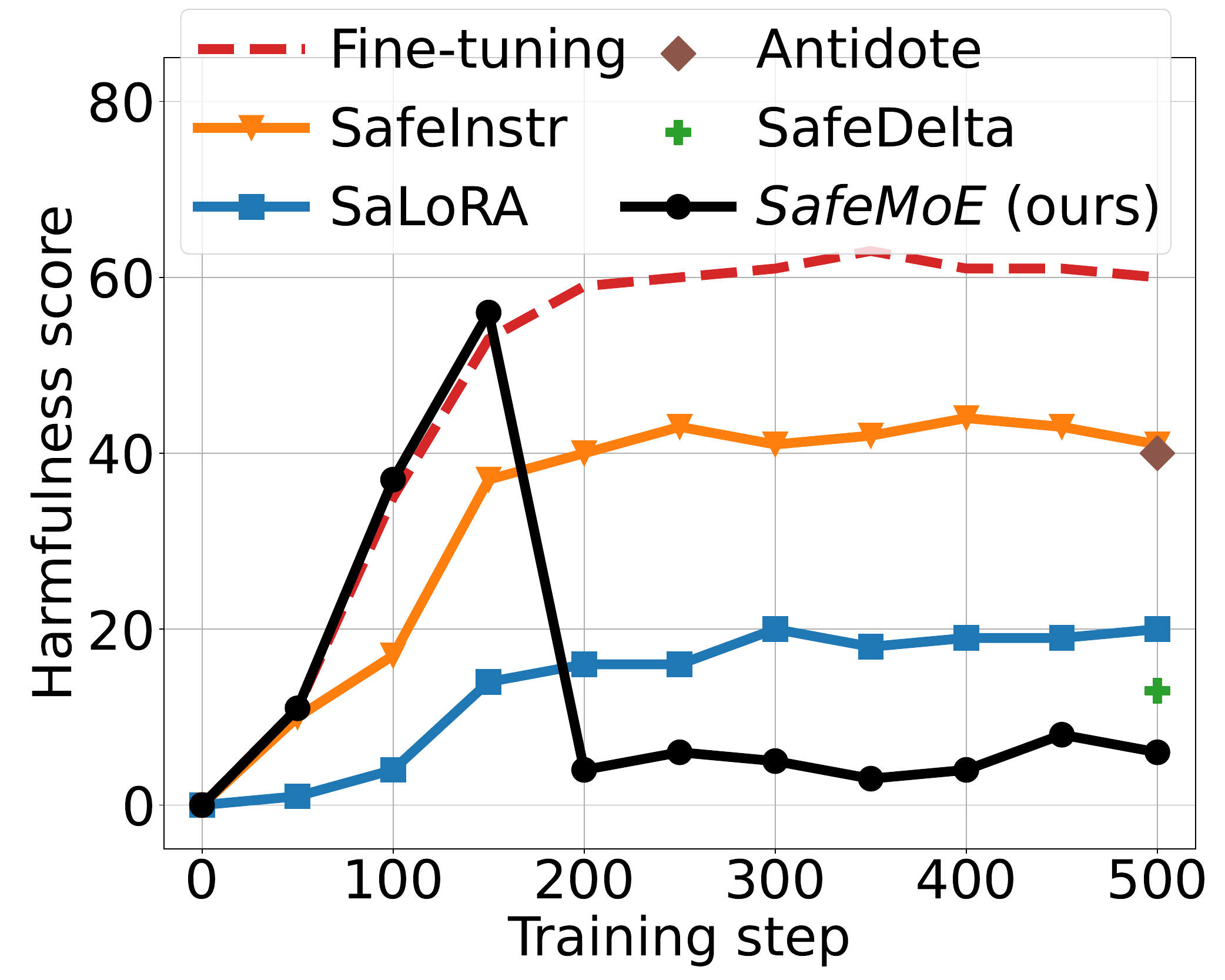}
        \caption{Harmfulness score}
    \end{subfigure}
    % \hspace{0.1cm}
    \begin{subfigure}[t]{0.495\linewidth}
        \centering
        \includegraphics[width=\linewidth]{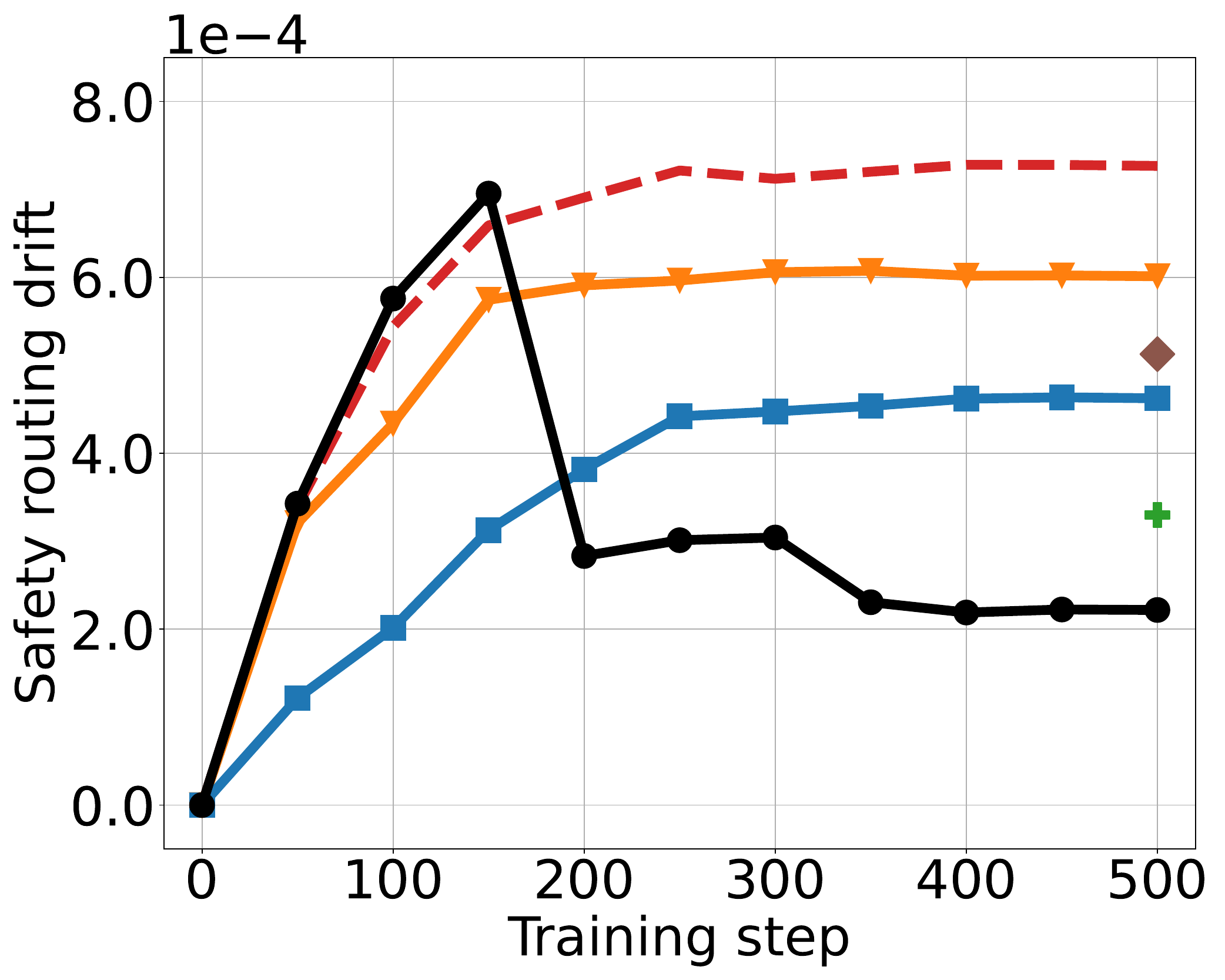}
        \caption{Safety routing drift}
    \end{subfigure}
    \caption{Effectiveness of defenses against HFT attacks.}
    \label{fig:dynamics_drift_baseline}
% \end{figure}
\end{wrapfigure}
However, existing defenses~\citep{huangbooster,li2025salora,lu2025safe} are primarily designed for the monolithic transformer architecture and overlook the superficial safety mechanisms of MoE LLMs, exhibiting limitations in mitigating fine-tuning risks.
In support of this, we conduct a preliminary experiment on the training dynamics of existing defenses under HFT attacks, including fine-tuning-stage methods~\citep{bianchi2024safety,li2025salora} and post-fine-tuning methods~\citep{huang2025antidote,lu2025safe}, as shown in Figure~\ref{fig:dynamics_drift_baseline}.
The safety routing drift is quantified as the deviation in routing weights on harmful instructions between the fine-tuned and safety-aligned models.
In MoE LLMs, these state-of-the-art defenses become less effective in preventing the safety routing drift and reducing harmfulness of fine-tuned models.

To address this limitation, we propose \ours{}, the first safe fine-tuning method tailored to safeguard MoE-based LLMs, which directly addresses the vulnerability in their safety mechanisms. Specifically, we design a regularization technique that aligns the routing decisions of a fine-tuned MoE LLM with those of the initial safety-aligned model by minimizing the KL-divergence of their routing weights on harmful inputs during fine-tuning. This encourages the fine-tuned model to direct harmful inputs to safety-critical experts as in the safety-aligned model. Accordingly, the fine-tuned model withstands the effects of HFT attacks while attaining comparable task utility. To reduce the overhead of \ours{}, we adopt a greedy optimization strategy that alternates between the fine-tuning and regularization steps rather than optimizing both simultaneously.

We conduct extensive experiments across eight widely used MoE LLMs, ranging from 7B to 141B parameters. Our safety evaluation results demonstrate the surpassing effectiveness and robustness of \ours{} in safeguarding MoE LLMs against fine-tuning risks. For example, it effectively mitigates an HFT attack that raises the harmfulness score of OLMoE (7B) to 62.0, reducing it to 5.0 with only a 1\% degradation in the task utility, outperforming state-of-the-art defenses significantly. This notable effectiveness remains consistent across diverse MoE LLMs, including larger and more advanced models such as gpt-oss and Llama 4, while maintaining their reasoning performance.

Based on an in-depth analysis of training dynamics, we confirm that our regularization technique is methodologically valid in preventing the safety routing drift and driving harmfulness reduction. We note that \ours{} is highly efficient, comparable to the baseline methods, with only approximately 2\% training time overhead in both LoRA and full fine-tuning, which demonstrates its practicality to large-scale MoE LLMs. Through these findings, we highlight the importance of architecture-aware designs of safe fine-tuning methods for MoE LLMs.

Our contributions are summarized as follows:
\begin{itemize}[leftmargin=30pt]
    \item We identify a vulnerability in the safety mechanism of MoE LLMs, where the drift in routing decisions for harmful inputs during fine-tuning undermines their safety.
    \item We propose \ours{}, an effective and efficient safe fine-tuning method tailored to MoE LLMs that preserves the routing decisions of the initial safety-aligned model.
    \item Through experiments on open-source MoE LLMs, we show that this vulnerability is consistent across diverse models and that \ours{} offers robust mitigation against HFT attacks.
\end{itemize}

\section{Preliminaries}

\begin{wrapfigure}{r}[0.7cm]{0.475\columnwidth}
% \vspace{-0.3cm}
    \includegraphics[width=\linewidth]{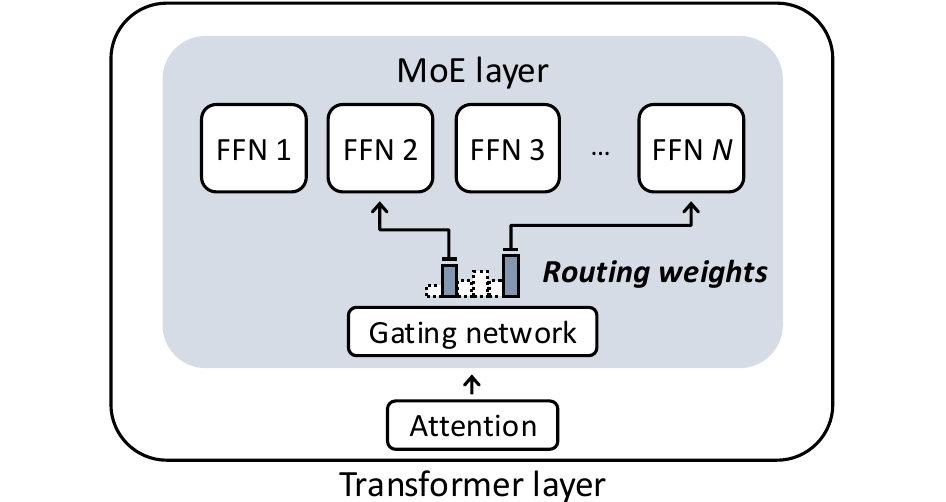}
    \caption{MoE LLM architecture.}
    \label{fig:moe_architecture}
\end{wrapfigure}

\textbf{Mixture-of-experts (MoE).} 
Figure~\ref{fig:moe_architecture} shows the standard MoE architecture for LLMs~\citep{shazeer2017outrageously,lepikhin2021gshard,du2022glam,komatsuzaki2023sparse,fedus2022switch}.
These MoE LLMs typically adhere to the transformer architecture, but the feed-forward network (FFN) layer in each transformer layer is replaced by an MoE layer.
The MoE layer consists of multiple independent FFNs, referred to as \textit{experts}, along with a gating network.
For each input token, the gating network dynamically assigns a \textit{routing weight} to every expert in the MoE layer based on the token's hidden state from the self-attention layer.
The top-$k$ experts are then selected for forward pass computation, and their outputs are combined according to the assigned weight
The MoE layer output is formalized as:
\begin{align}
    h_{\text{MoE}} = \sum_{i \in N} \texttt{TopK}(\sigma(\vr(x))) \text{FFN}_i(x),
\end{align}
where $N$ is the number of experts, $\sigma$ is Softmax, $\vr(h) \in \mathbb{R}^N$ is routing weights on an input token $x$, and $\text{FFN}_i$ is the $i$-th expert layer.
By leveraging this conditional computation approach, the MoE architecture provides substantial efficiency gains in both training and inference, which has driven its adoption in recent advanced LLMs.
%
% \sooelnote{I recommend adding a simple equation for routing weight vectors and MoE LLM weight so that readers can easily follow Definition 3.1 later and how this weight vector can affect token outputs.}
%

% \sooelnote{I am not sure what the following means.}
% Several MoE LLMs incorporate shared experts that are always activated to capture general knowledge~\citep{}.

\textbf{Superficial safety mechanism in MoE.} 
Routing decisions in MoE LLMs play a crucial role in composing the outputs and have a direct impact on their safety.
Recent studies~\citep{lai2025safex,fayyaz2025steering} show that harmful instructions to safety-aligned MoE LLMs, such as \textit{``How to make a bomb?''}, consistently trigger specific experts, referred to \textit{safety-critical experts}.
Moreover, masking these experts causes significant degradation in safety even when the model parameters remain unchanged.
This indicates that the safety alignment of MoE LLMs heavily depends on routing decisions that determine the activation of safety-critical experts.

\textbf{Harmful fine-tuning (HFT) attacks.} 
Safety alignment has become an essential step in ensuring that LLM outputs are harmless and aligned with human values~\citep{ouyang2022training,rafailov2023direct}.
However, researchers show that this alignment can be undone through user fine-tuning~\citep{qi2024fine,yang2024shadow,zhan2024removing}. Specifically, they find that HFT attacks that inject a small portion of harmful samples into the training dataset, or even benign fine-tuning alone, can severely impair LLM safety.
Considering the increasing availability of fine-tuning API services~\citep{openaifinetuning,geminifinetuning}, HFT attacks have become a practical threat to LLM providers.

To mitigate this threat, several methods have been proposed to enhance the safety of LLM fine-tuning---for example, augmenting datasets with safe samples~\citep{bianchi2024safety,zong2024safety}, constraining training to prevent harmful-direction drift~\citep{huang2024lazy,wu2025mitigating}, and pruning harmful parameters from the fine-tuned models~\citep{huang2025antidote,lu2025safe}.

However, existing defenses focus solely on monolithic LLMs and overlook the distinct architecture of MoE LLMs based on dynamic input routing and their superficial safety mechanisms. To the best of our knowledge, no prior work has investigated safe fine-tuning strategies tailored to MoE LLMs.

\section{Safety Vulnerability in MoE LLMs}
\label{sec:safety_analysis}

% Existing related studies have mainly focused on the vulnerability of safety-aligned monolithic LLMs to harmful fine-tuning (HFT) attacks~\citep{qi2024fine,yang2024shadow,zhan2024removing,bianchi2024safety,zong2024safety,huang2024lazy,li2025salora}.
% %
% However, safety problems of MoE-based LLMs under these practical threat scenarios remain largely unexplored.
% %
% Therefore, we aim to identify the safety vulnerability of the MoE LLM architecture by analyzing its training mechanisms, then ultimately propose a fine-tuning solution tailored for safeguarding MoE LLMs.

Given that the safety of MoE LLMs depends on safety-critical experts~\citep{lai2025safex,fayyaz2025steering}, we posit that \textit{safety degradation in fine-tuned MoE LLMs arises from substantial deviations in routing decisions for harmful instructions compared to those in the initial safety-aligned models}.

To verify this, we first define \textit{safety routing drift} for harmful inputs. This metric quantifies a difference between the routing weights of a safety-aligned MoE LLM and its fine-tuned counterpart:
\begin{definition}[Safety routing drift]
    Let $\vw_{align}$ be a safety-aligned MoE LLM.
    Given a fine-tuned model $\vw_{ft}$, the safety routing drift for a harmful instruction input $x$ is defined as:
    \begin{align}
        d(\vw_{ft}, x) = D_{KL}\left( \sigma(\vr({x|\vw_{align}})) \, \big|\big| \, \sigma(\vr({x|\vw_{ft}})) \right),
    \end{align}
    where $D_{KL}(\cdot || \cdot)$ denotes KL divergence, $\sigma$ is Softmax, and $\vr({\cdot|\vw}) \in \mathbb{R}^N$ denotes the routing weight vector over $N$ experts of model $\vw$ for an input.
    \label{def:safety_routing_drift}
\end{definition}

We then analyze the training dynamics of three MoE LLMs, including OLMoE~\citep{muennighoff2025olmoe}, Qwen1.5 MoE~\citep{qwen1.5moe}, and DeepSeek V2~\citep{liu2024deepseek}, and measure the correlation between the safety routing drift and harmfulness of their fine-tuned models.
Specifically, we consider two fine-tuning scenarios: i) benign fine-tuning on 5.5k samples from the Alpaca dataset~\citep{alpaca}, and ii) HFT on a combined dataset of 5k task-specific samples from SAMSum~\citep{gliwa-etal-2019-samsum} and 500 harmful samples from BeaverTails~\citep{ji2023beavertails}.
For each fine-tuned LLM, we compute safety routing drift on the last tokens of harmful instructions from JailbreakBench~\citep{chao2024jailbreakbench}. Harmfulness scores are computed as the proportion of \textit{unsafe} responses, evaluated using Llama-Guard-4-12B~\citep{meta2025llamaguard4}.

\begin{figure}[t]
    \centering
    \begin{subfigure}[t]{0.327\linewidth}
        \centering
        \includegraphics[width=\linewidth]{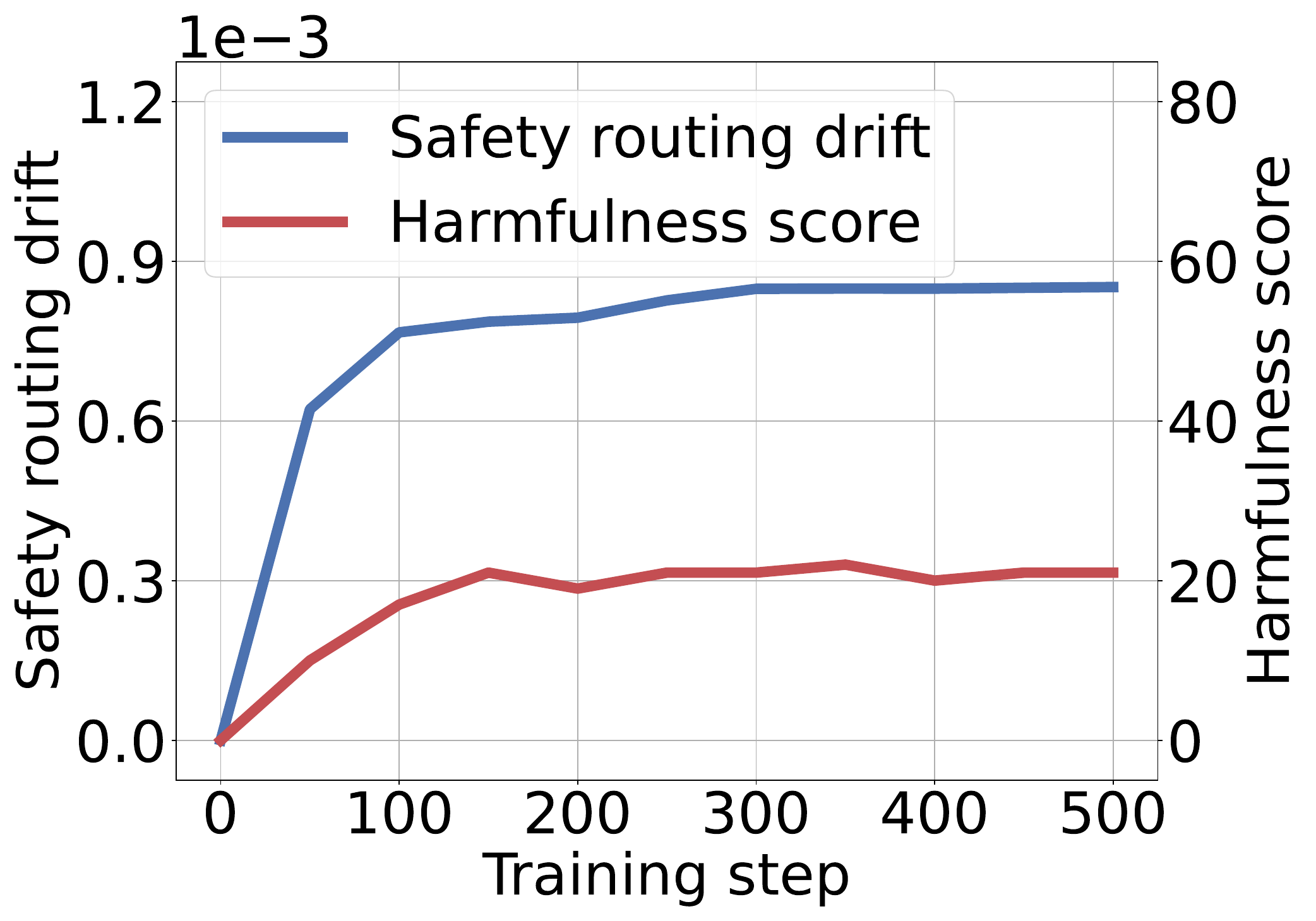}
        \caption{OLMoE (Benign FT)\\($r=0.9823$, $p=7e{-8}$)}
    \end{subfigure}
    % \hspace{0.1cm}
    \begin{subfigure}[t]{0.327\linewidth}
        \centering
        \includegraphics[width=\linewidth]{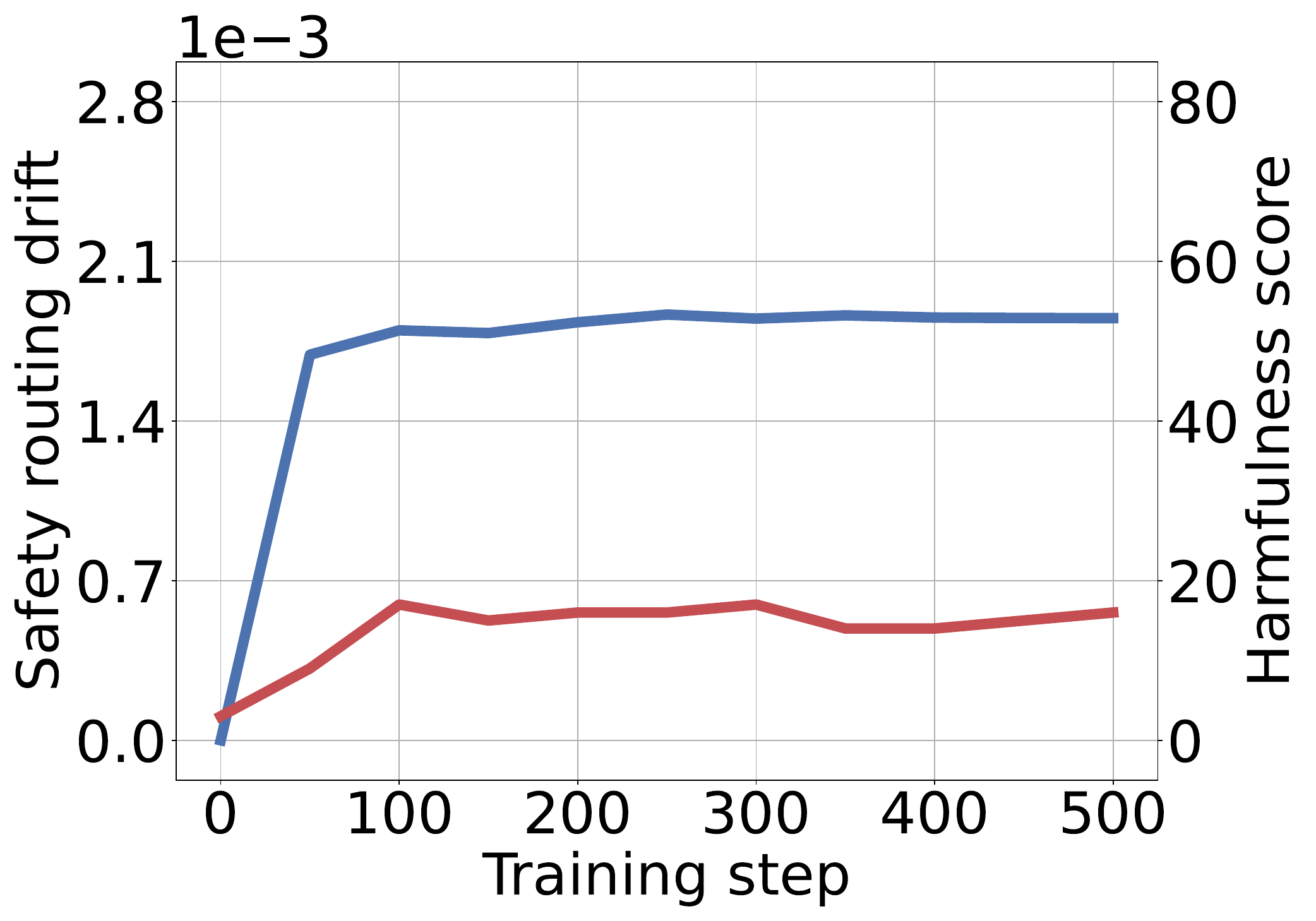}
        \caption{Qwen1.5 MoE (Benign FT)\\($r=0.8977$, $p=1e{-4}$)}
    \end{subfigure}
    % \hspace{0.1cm}
    \begin{subfigure}[t]{0.327\linewidth}
        \centering
        \includegraphics[width=\linewidth]{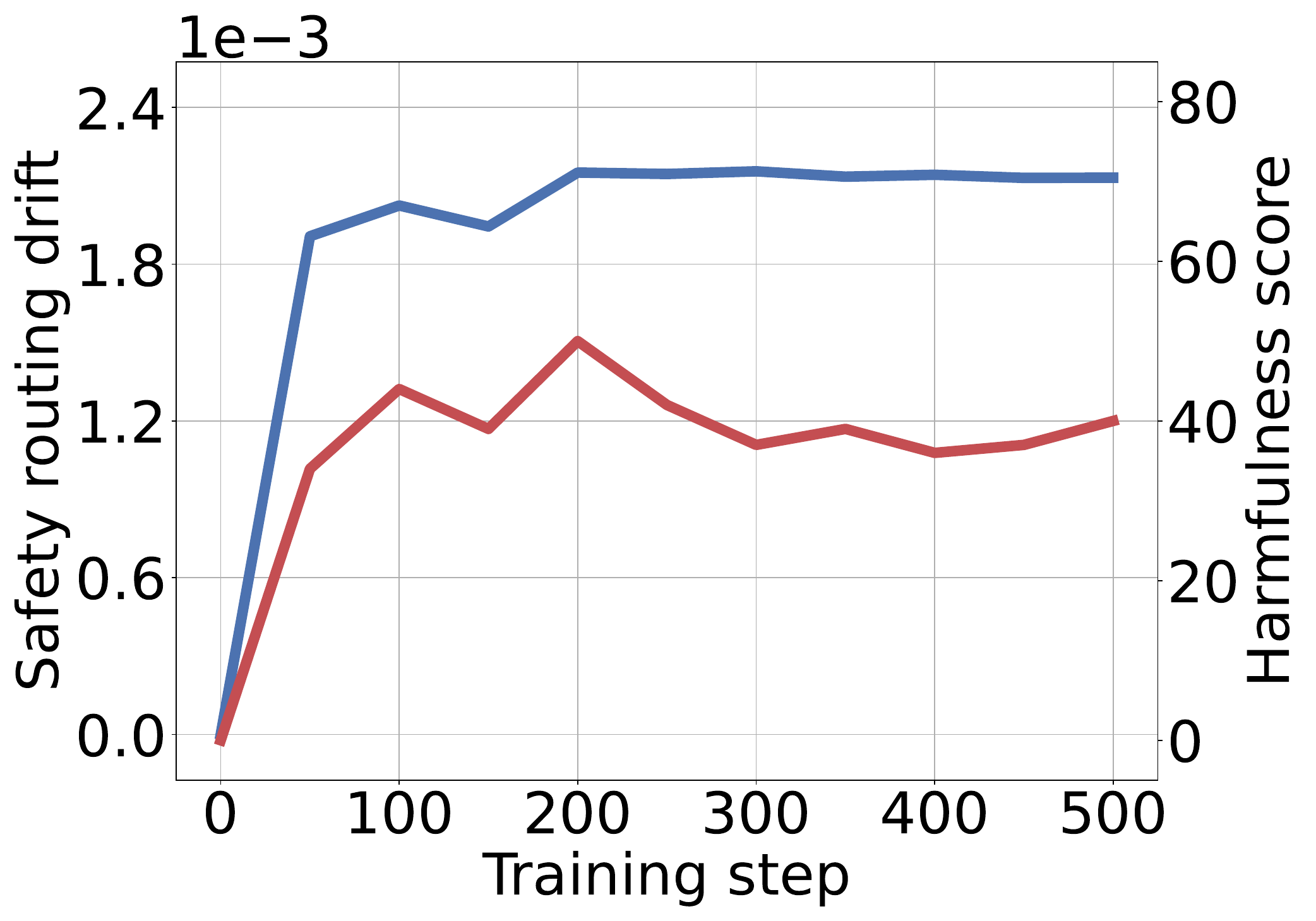}
        \caption{DeepSeek V2 (Benign FT)\\($r=0.9354$, $p=2e{-5}$)}
    \end{subfigure}
    \begin{subfigure}[t]{0.327\linewidth}
        \centering
        \includegraphics[width=\linewidth]{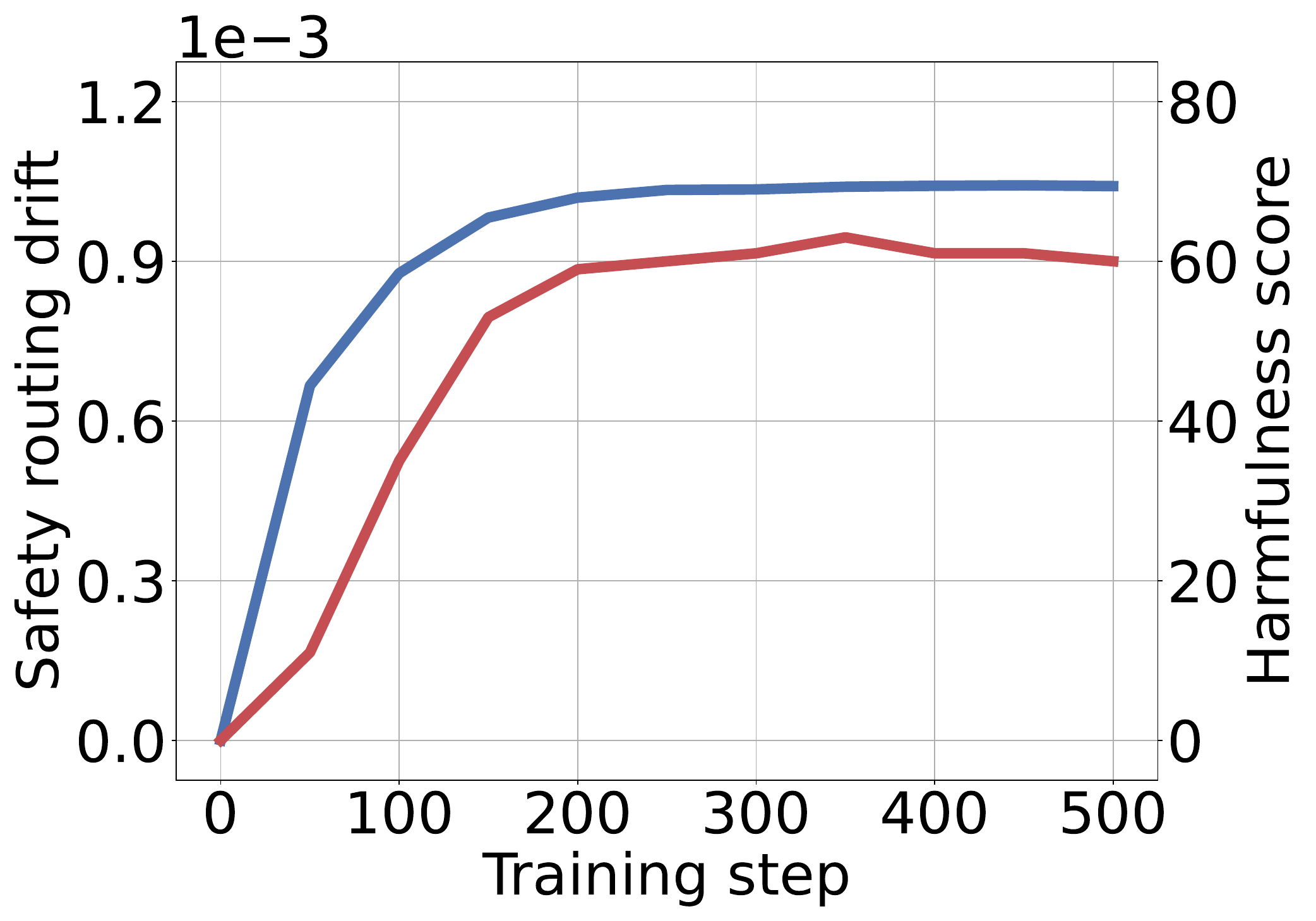}
        \caption{OLMoE (HFT attack)\\($r=0.9681$, $p=1e{-6}$)}
    \end{subfigure}
    % \hspace{0.02cm}
    \begin{subfigure}[t]{0.327\linewidth}
        \centering
        \includegraphics[width=\linewidth]{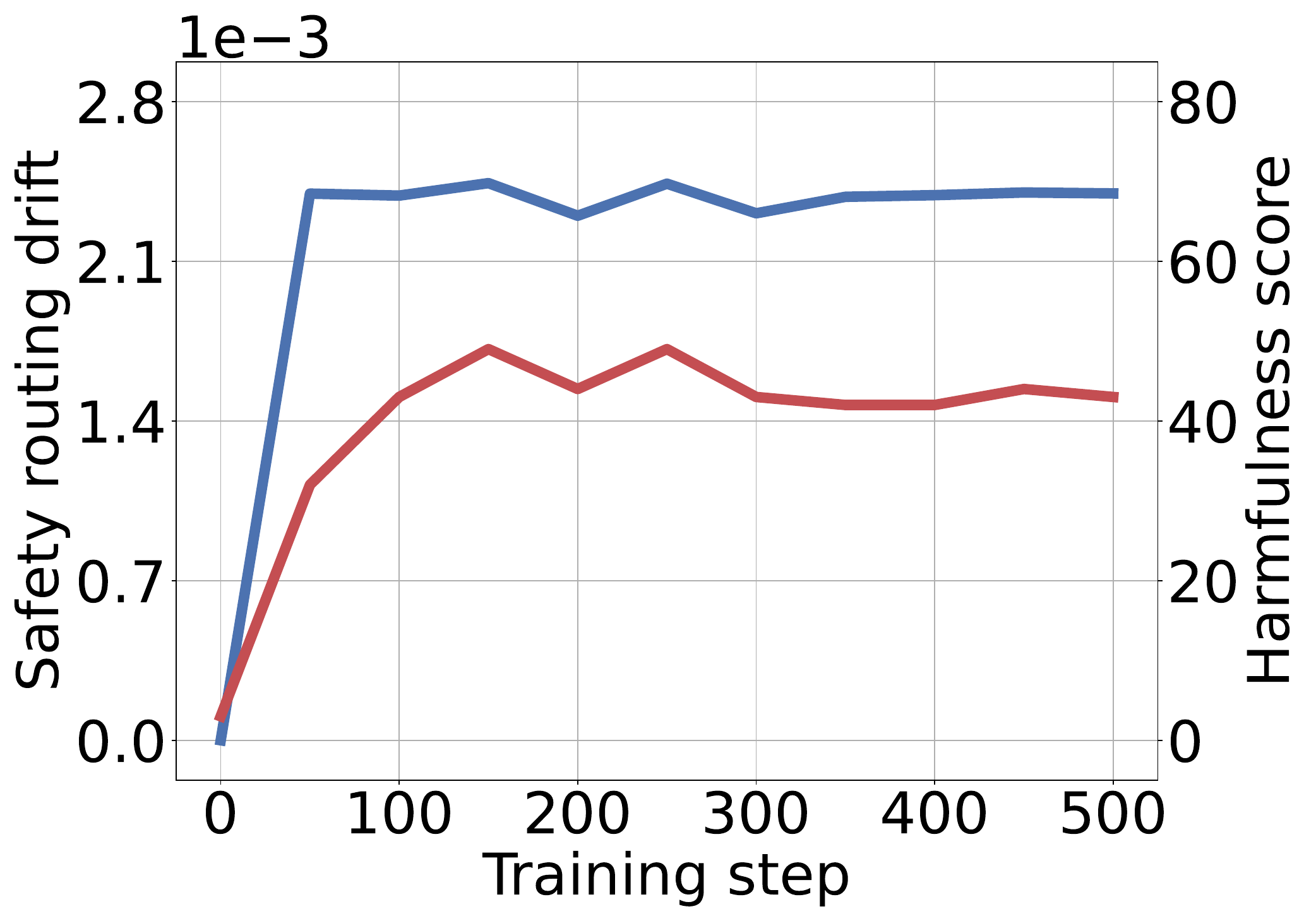}
        \caption{Qwen1.5 MoE (HFT attack)\\($r=0.9445$, $p=1e{-5}$)}
    \end{subfigure}
    % \hspace{0.02cm}
    \begin{subfigure}[t]{0.327\linewidth}
        \centering
        \includegraphics[width=\linewidth]{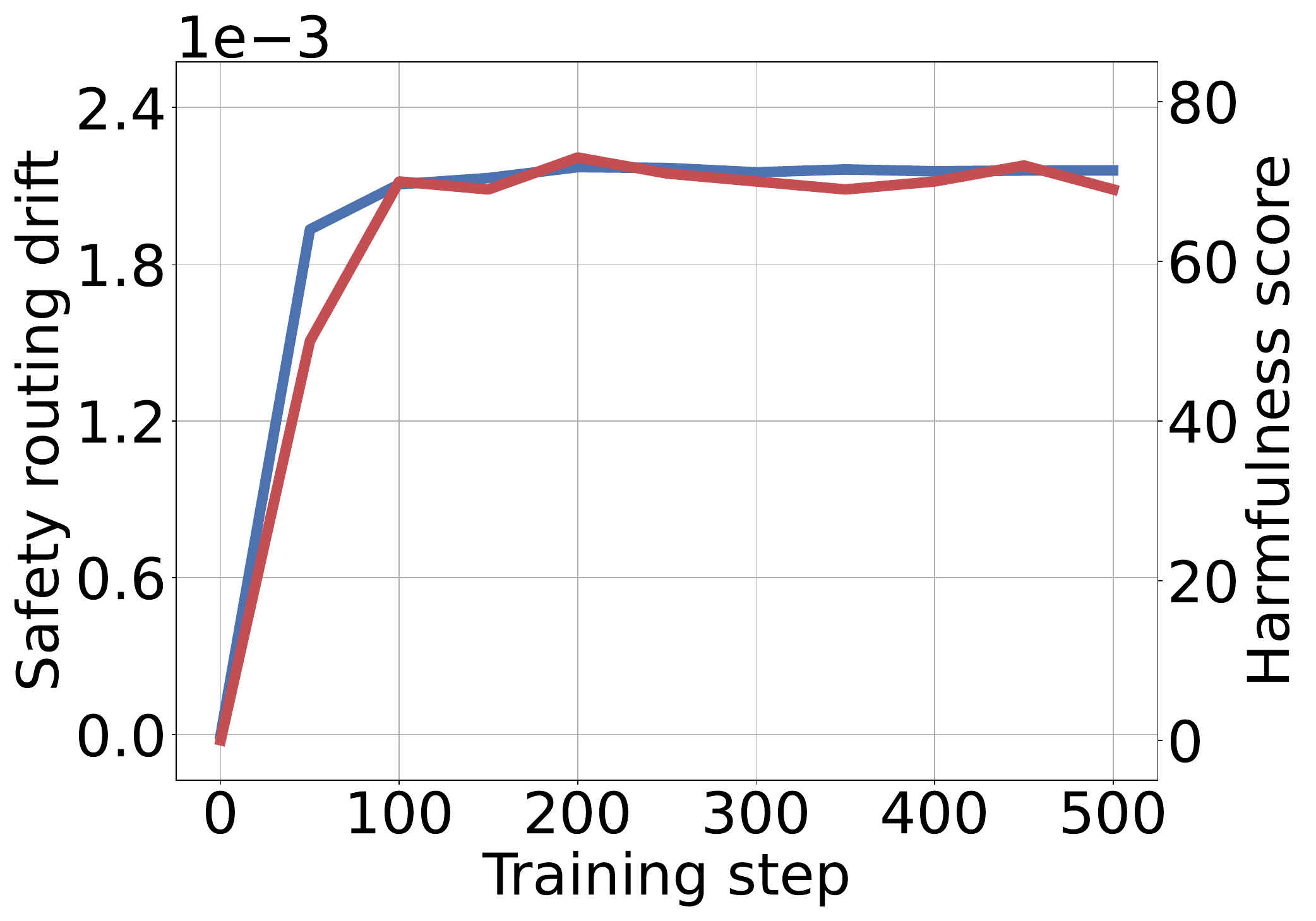}
        \caption{DeepSeek V2 (HFT attack)\\($r=0.9933$, $p=1e{-10}$)}
    \end{subfigure}
    \caption{Safety routing drift and harmfulness of MoE LLMs over training steps. Results of t-tests for Pearson correlation coefficients are reported ($r$: correlation coefficient, $p$: $p$-value).}
    \label{fig:routing_drift}
\end{figure}

Figure~\ref{fig:routing_drift} presents the results of our analysis.
In both fine-tuning scenarios, we observe significant safety routing drift from the initial safety-aligned model, with the drift metrics increasing as training progresses.
Notably, routing decisions for harmful instructions drift even under benign fine-tuning, causing nontrivial increases in harmfulness. This demonstrates that the superficial safety mechanism of MoE LLMs is highly fragile and easily disrupted by fine-tuning. Moreover, the magnitude of safety routing drift is strongly correlated with harmfulness scores across the models, with high statistical significance ($p \ll 0.05$). Harmful fine-tuning (HFT) attacks even further amplify this drift, producing substantially larger increases in harmfulness than benign fine-tuning.
%%Importantly, routing decisions for harmful instructions drift even under benign fine-tuning unintentionally, leading to nontrivial increases in harmfulness. This indicates that the superficial safety mechanism of MoE LLMs is highly susceptible to fine-tuning procedures.
%
%%Moreover, the extent of safety routing drift is strongly correlated with the harmfulness score of the MoE LLM with high statistical significance ($p$-value $\ll$ 0.05).

\noindent\textbf{Motivation for an MoE-specific defense.} Our analysis shows that the safety routing drift is highly correlated with the safety degradation in fine-tuned MoE LLMs.
This suggests the need for an approach that preserves the initial routing decisions of the safety-aligned models on harmful inputs to mitigate fine-tuning risks in MoE LLMs.
However, prior defenses, designed for monolithic LLMs, are limited in preventing the safety routing drift and reducing the harmfulness of fine-tuned MoE LLMs, as shown in our preliminary experiments (see Figure~\ref{fig:dynamics_drift_baseline}).
These findings motivate a new defense that reflects the unique architecture of MoE LLMs and their safety mechanism.

\section{Safe MoE Fine-Tuning Method}
\label{sec:safe_fine_tuning}

\begin{figure}[t]
    \centering
    \includegraphics[width=0.99\linewidth]{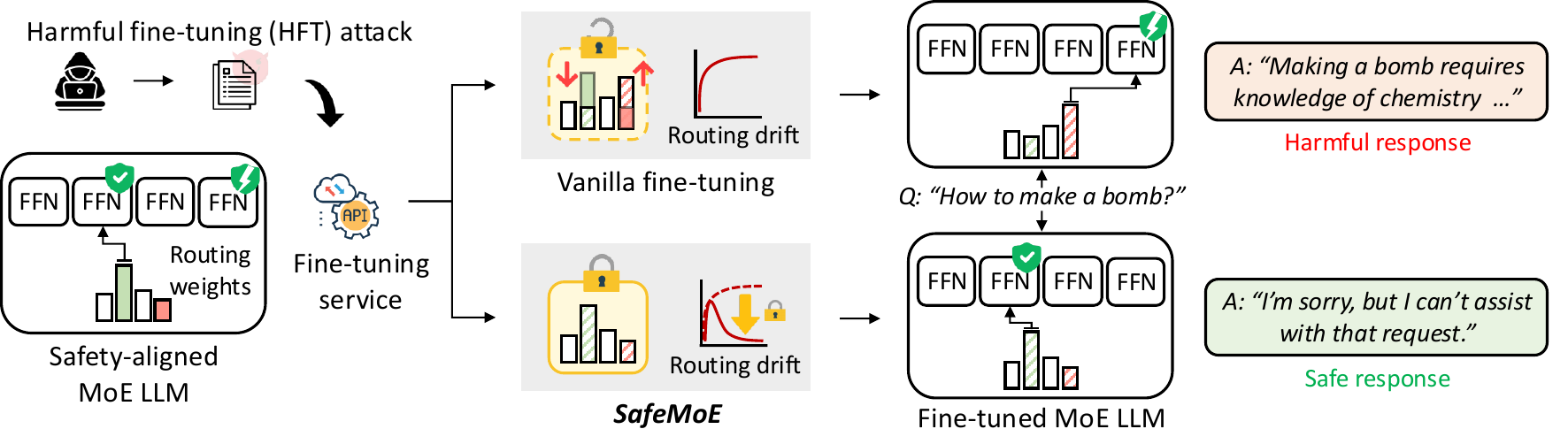}
    \caption{Overview of \ours{}. It mitigates the safety routing drift by directly constraining this drift during fine-tuning, thereby effectively safeguarding MoE LLMs against HFT attacks.}
    \label{fig:overview}
\end{figure}

\textbf{Safety routing drift regularization.} 
%
%Based on our analysis identifying safety routing drift as a key vulnerability in MoE LLMs, 
%
We propose a novel fine-tuning approach designed to improve the safety of fine-tuned MoE LLMs, as illustrated in Figure~\ref{fig:overview}.
Specifically, during fine-tuning, we propose leveraging a regularization objective that constrains the safety routing drift (Definition~\ref{def:safety_routing_drift}).
This objective penalizes deviations in the routing weight distributions of an MoE LLM $\vw$ under fine-tuning from those of the initial safety-aligned model $\vw_{align}$ on harmful instructions:
\begin{align}
    \mathcal{L}_{reg} (\vw) = \mathbb{E}_{x \in \mathcal{D}_{h}} \mathbb{E}_{l \in L} D_{\mathrm{KL}}\left( \sigma (\vr^{(l)}({x|\vw_{align}})/\tau) \, \big|\big| \, \sigma (\vr^{(l)}({x|\vw})/\tau) \right),
    \label{eq:regularization}
\end{align}
where $\mathcal{D}_{h}$ is the harmful instruction dataset, and $L$ is the set of transformer layers.
We apply regularization to the routing weights assigned to the last token of each harmful instruction.
The temperature $\tau$ controls the strength of regularization. A smaller $\tau$ (e.g., $<1.0$) further enhances safety by sharpening the routing weight distribution, which enables the regularization to focus more on top-ranked safety-critical experts and tightly constrains the routing weights of those experts.

\textbf{Bi-level greedy optimization}. We integrate the regularization into supervised fine-tuning on a task-specific dataset $\mathcal{D}_{ft}$ to solve:
\begin{align}
    \mathop{\arg\min}_{\vw} \ \mathcal{L}_{\mathrm{sft}}(\vw) + \mathcal{L}_{\mathrm{reg}} (\vw).
\end{align}
Simultaneously optimizing both losses at every training step, however, incurs substantial computational overhead.
To address this, we reframe the joint optimization into a bi-level greedy approach that alternates between the supervised fine-tuning and regularization steps. 

The greedy optimization process is described in Algorithm~\ref{alg:greedy}.
We first precompute the routing weights of the initial safety-aligned MoE LLM over all harmful instructions in $\mathcal{D}_h$ to avoid redundant forward passes in the regularization steps (line 2).
During training with the fine-tuning loss $\mathcal{L}_{sft}$, we insert safety-preserving steps using the regularization loss $\mathcal{L}_{reg}$ every $T_{reg}$ steps (line 6).

\section{Experiments}

\subsection{Setup}

\textbf{MoE LLMs.} 
We conduct experiments on eight widely adopted MoE LLMs: OLMoE-1B-7B-Instruct~\citep{muennighoff2025olmoe}, Qwen1.5-MoE-A2.7B-Chat~\citep{qwen1.5moe}, and DeepSeek-V2-Lite-Chat~\citep{liu2024deepseek}, gpt-oss-20b~\citep{openai2025gptoss}, Qwen3-30B-A3B~\citep{qwen3technicalreport}, Phi-3.5-MoE-Instruct~\citep{phi3.5moe}, and two models with on-the-fly 4-bit quantization, Llama-4-Scout-17B-16E-Instruct~\citep{meta2025llama4} and Mixtral-8x22B-Instruct-v0.1~\citep{mistral2025mixtral}.
All models used are safety-aligned versions.
%The number of experts ranges from 16 to 128, covering diverse fine-tuning scenarios and settings.
We describe model details in the Appendix~\ref{appendix:exprimental_setting}.

\textbf{Fine-tuning datasets.} We consider two tasks: SAMSum~\citep{gliwa-etal-2019-samsum} for dialogue summarization and SQL~\citep{sqldataset} for SQL query generation, both widely adopted for implementing HFT attack scenarios~\citep{yang2025alleviating,wang2024backdooralign}.
For effective attacks, we conduct supervised fine-tuning with 5k task samples combined with 500 harmful samples from BeaverTails~\citep{ji2023beavertails}.
Fine-tuning is performed with LoRA~\citep{hu2022lora}, as detailed in Appendix~\ref{appendix:exprimental_setting}.

\textbf{Metrics.} We reports two metrics: \textit{Fine-tuning accuracy (FA)} and \textit{Harmfulness score (HS)}. FA measures task utility---Rouge-1 score for SAMSum and exact match accuracy for SQL.
To assess reasoning performance, we measure accuracy on 570 samples (10 from each categories) from MMLU-Redux-2.0~\citep{gema2025we}.
HS refers to the proportion of responses to JailbreakBench~\citep{chao2024jailbreakbench} instructions that are classified as \textit{unsafe} by Llama-Guard-4-12B~\citep{meta2025llamaguard4}.

\textbf{Baselines.}
We evaluate four state-of-the-art defenses against HFT attacks.
Among fine-tuning-stage methods, SafeInstr~\citep{bianchi2024safety} augments fine-tuning datasets with additional safe samples, while SaLoRA~\citep{li2025salora} initializes LoRA layers with weights optimized on safe samples.
We also consider two post-fine-tuning weight modification approaches.
Antidote~\citep{huang2025antidote} prunes harmful parameters by analyzing their contribution to harmful instructions.
SafeDelta~\citep{lu2025safe} selects delta parameters that maximize utility while minimizing safety degradation.
%
% For all baselines, we tune hyperparameters and report results optimized for a balance between model safety and task utility.

\textbf{\ours{} implementation.} We use 100 harmful instruction samples from SafeInstr~\citep{bianchi2024safety} as the dataset $\mathcal{D}_{h}$.
We select $\tau$ as the smallest value in the range 0.1-1.3 that allows a 1\% degradation in fine-tuning accuracy.
By default, $T_{\mathrm{reg}}$ is set to the number of steps per epoch.

\subsection{Safety Evaluation}

Table~\ref{tab:eval_main} reports the defense effectiveness of \ours{} on three widely used MoE LLMs.
The initial safety-aligned models prior to fine-tuning (first row) are highly safe, exhibiting nearly zero harmfulness scores.
Vanilla fine-tuning substantially undermines safety while improving task performance (second row).
\ours{} attains the lowest harmfulness score while incurring only minimal loss in fine-tuning accuracy.
For example, it reduces the harmfulness score of DeepSeek V2 fine-tuned on the SQL task from 72.0 to 4.0.
\ours{}'s superior effectiveness is attributed to its MoE-specific design, which explicitly encourages the routing of harmful instructions to safety-critical experts.

\begin{table}[t]
    \centering
    \caption{Safety evaluation of MoE LLMs. We report fine-tuning accuracy (FA$\uparrow$) and harmfulness score (HS$\downarrow$) for the SAMSum and SQL tasks. The number of parameters is denoted as (active/total).}
    \small
    \begin{tabular} {@{ } l @{\quad} c c c c @{ } c c c c c c @{} c c c c @{ }}
        \toprule
        \multirow{3.5}{*}{Method} & \multicolumn{4}{c}{\textbf{OLMoE (1.3B/6.9B)}} && \multicolumn{4}{c}{\textbf{Qwen1.5 MoE (2.7B/14.3B)}} && \multicolumn{4}{c}{\textbf{DeepSeek V2 (2.4B/15.7B)}} \\ \cmidrule{2-5}\cmidrule{7-10}\cmidrule{12-15}
        & \multicolumn{2}{c}{SAMSum} & \multicolumn{2}{c}{SQL} && \multicolumn{2}{c}{SAMSum} & \multicolumn{2}{c}{SQL} && \multicolumn{2}{c}{SAMSum} & \multicolumn{2}{c}{SQL} \\ 
        & FA & HS & FA & HS && FA & HS & FA & HS && FA & HS & FA & HS \\ \midrule
        Aligned & 31.8 & 0 & 43.0 & 0 && 36.6 & 3.0 & 38.4 & 3.0 && 34.0 & 0 & 29.6 & 0 \\
        Fine-tuning & 49.3 & 62.0 & 58.5 & 64.0 && 50.4 & 49.0 & 70.2 & 37.0 && 52.0 & 70.0 & 70.1 & 72.0  \\ \midrule
        SafeInstr & \textbf{49.5} & 46.0 & \underline{58.9} & 41.0 && \underline{50.6} & \underline{11.0} & \underline{69.3} & \underline{8.0} && \textbf{52.1} & \underline{25.0} & \textbf{70.2} & \underline{21.0} \\
        % SAP & 48.7 & 32.0 &  & \\
        SaLoRA & \underline{48.9} & 24.0 & 54.5 & 40.0 && 48.9 & 28.0 & 54.9 & 33.0 && 50.1 & 66.0 & 62.0 & 74.0 \\
        Antidote & 48.7 & 40.0 & 57.5 & 44.0 && 49.3 & 18.0 & 68.6 & 29.0 && 50.7 & 70.0 & 65.3 & 62.0 \\ 
        SafeDelta & 48.6 & \underline{13.0} & 57.4 & \underline{33.0} && 50.2 & 22.0 & 69.2 & 30.0 && 51.0 & 47.0 & 69.0 & 72.0\\\midrule
        \ours{} & 48.9 & \textbf{5.0} & \textbf{59.0} & \textbf{17.0} && \textbf{50.6} & \textbf{0} & \textbf{69.5} & \textbf{1.0} && \underline{51.0} & \textbf{1.0} & \underline{69.1} & \textbf{4.0} \\ \bottomrule
    \end{tabular}%
    \label{tab:eval_main}
\end{table}

\begin{table}[t]
    \centering
    \caption{Extended safety evaluation on five large-scale MoE LLMs. We report harmfulness score (HS$\downarrow$) and reasoning performance (MMLU$\uparrow$) using MMLU-Redux-2.0.}
    \small
    % \resizebox{\linewidth}{!}{%
        \begin{tabular} {l @{ } c c @{ } c c c @{ } c c c @{ } c c c @{ } c c c}
            \toprule 
            \multirow{2.5}{*}{Method} & \multicolumn{2}{c}{\makecell{\textbf{gpt-oss}\\\textbf{(3.6B/20.9B)}}} & &\multicolumn{2}{c}{\makecell{\textbf{Qwen3 MoE}\\\textbf{(3.3B/30.5B)}}} & & \multicolumn{2}{c}{\makecell{\textbf{Phi 3.5 MoE}\\\textbf{(6.6B/41.9B)}}} & & \multicolumn{2}{c}{\makecell{\textbf{Llama 4}\\\textbf{(17B/109B)}}} & & \multicolumn{2}{c}{\makecell{\textbf{Mixtral}\\\textbf{(39B/141B)}}} \\ \cmidrule{2-3}\cmidrule{5-6}\cmidrule{8-9}\cmidrule{11-12}\cmidrule{14-15}
            & MMLU & HS & & MMLU & HS & & MMLU & HS & & MMLU & HS & & MMLU & HS \\ \midrule
            Aligned & 85.4 & 2.0 && 89.6 & 1.0 && 83.3 & 2.0 && 90.4 & 7.0 && 78.9 & 7.0 \\
            Fine-tuning & 77.5 & 84.0  & & 89.1 & 67.0 & & 80.7 & 83.0 & & 89.5 &  79.0 && 66.5 & 78.0 \\ \midrule
            \ours{} & 79.6 & 7.0  & & 88.8 & 4.0 & & 81.4 & 2.0 & & 89.8 & 3.0 && 78.4 & 8.0\\
            \bottomrule
        \end{tabular}%
    % }
    \label{tab:eval_large}
\end{table}

% gpt base 2.0 85.43
% gpt def 7.0 79.6

%qwen 1.0 89.6

% llama base 7.0 90.35

% mixtral base 7.0 78.9
% mixtral ft 78.0 66.5
% mixtral def 8.0 78.4

In contrast, baseline methods fail to effectively mitigate safety degradation. SafeInstr, which adopts an architecture-agnostic strategy based on safety data augmentation, achieves moderate defense performance but still leaves substantial harmfulness.
SaLoRA and post-fine-tuning methods (Antidote and SafeDelta) assume all parameters are always activated, as in monolithic LLMs. In MoE LLMs, however, the dynamically changing active parameters hinder full optimization of these methods.
Furthermore, they cause significant degradation in fine-tuning accuracy with only marginal harmfulness reduction, even though we extensively tune their hyperparameters (see Appendix~\ref{appendix:baseline_tuning}).

These results highlight the importance of architecture-aware defenses that directly address vulnerabilities in the safety mechanism. In Qwen1.5-MoE, we observe slightly improved safety compared to the safety-aligned models. This effect appears to result from the generalization ability of \ours{}, which promotes stronger activation of top-ranked safety-critical experts (see Appendix~\ref{appendix:activation}).

\noindent\textbf{Results on larger MoE LLMs.} We further evaluate \ours{} on recent larger MoE LLMs under a strong HFT attack scenario using 500 purely harmful samples~\citep{bianchi2024safety,lu2025safe,hsu2024safe}, as shown in Table~\ref{tab:eval_large}. These models employ diverse MoE configurations (e.g, activating 1 expert among 16 in Llama 4 and activating 8 among 128 in Qwen3 MoE) and distinct reasoning strategies (e.g., multi-level reasoning in gpt-oss and the thinking mode in Qwen3 MoE). Despite their differences, \ours{} generally achieves strong defense performance while effectively preserving the models' reasoning capability. For gpt-oss, Phi 3.5 MoE, and Mixtral, \ours{} even alleviates the degradation of reasoning performance observed in vanilla fine-tuning by preventing excessive overfitting to the HFT attack data. Overall, these results demonstrate the practicality of 
\ours{} for real-world fine-tuning services with high-capable MoE LLMs.

\begin{figure}[t]
    \centering
    \begin{subfigure}[t]{0.327\linewidth}
        \centering
        \includegraphics[width=\linewidth]{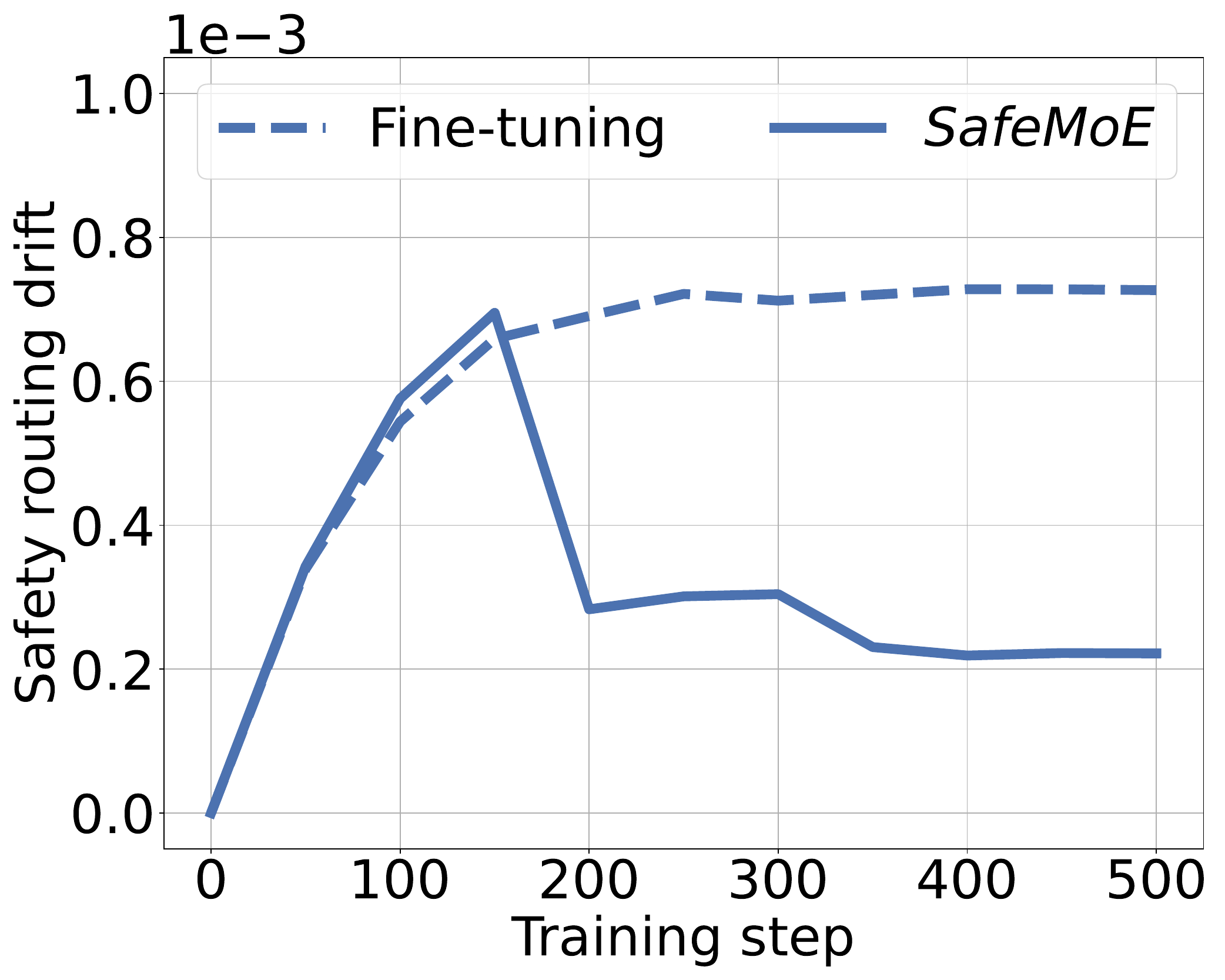}
        \caption{Safety routing drift}
        \label{subfig:dynamics_drift_safemoe}
    \end{subfigure}
    % \hspace{0.3cm}
    \begin{subfigure}[t]{0.327\linewidth}
        \centering
        \includegraphics[width=\linewidth]{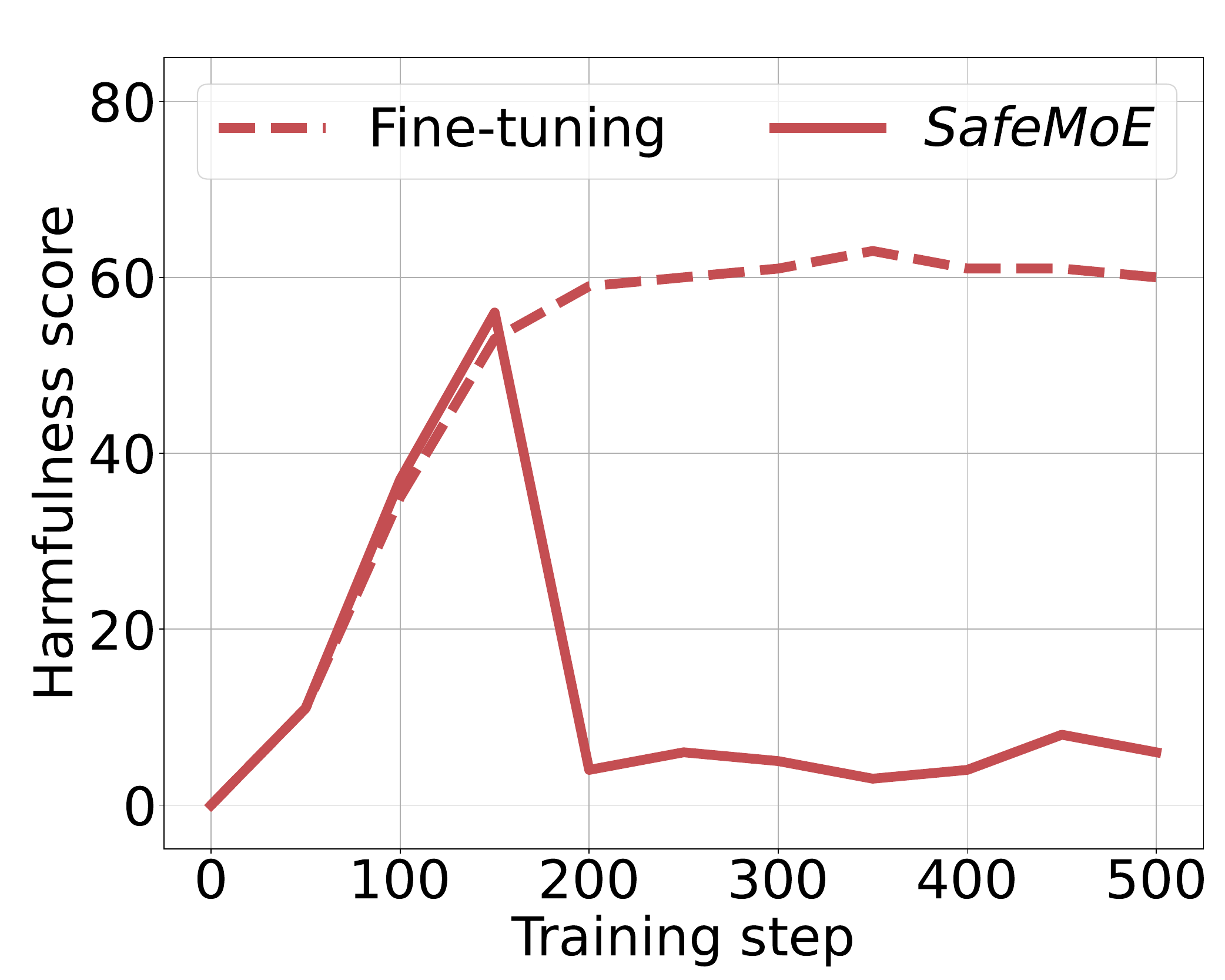}
        \caption{Harmfulness score}
        \label{subfig:dynamics_hs_safemoe}
    \end{subfigure}
    % \hspace{0.3cm}
    \begin{subfigure}[t]{0.327\linewidth}
        \centering
        \includegraphics[width=\linewidth]{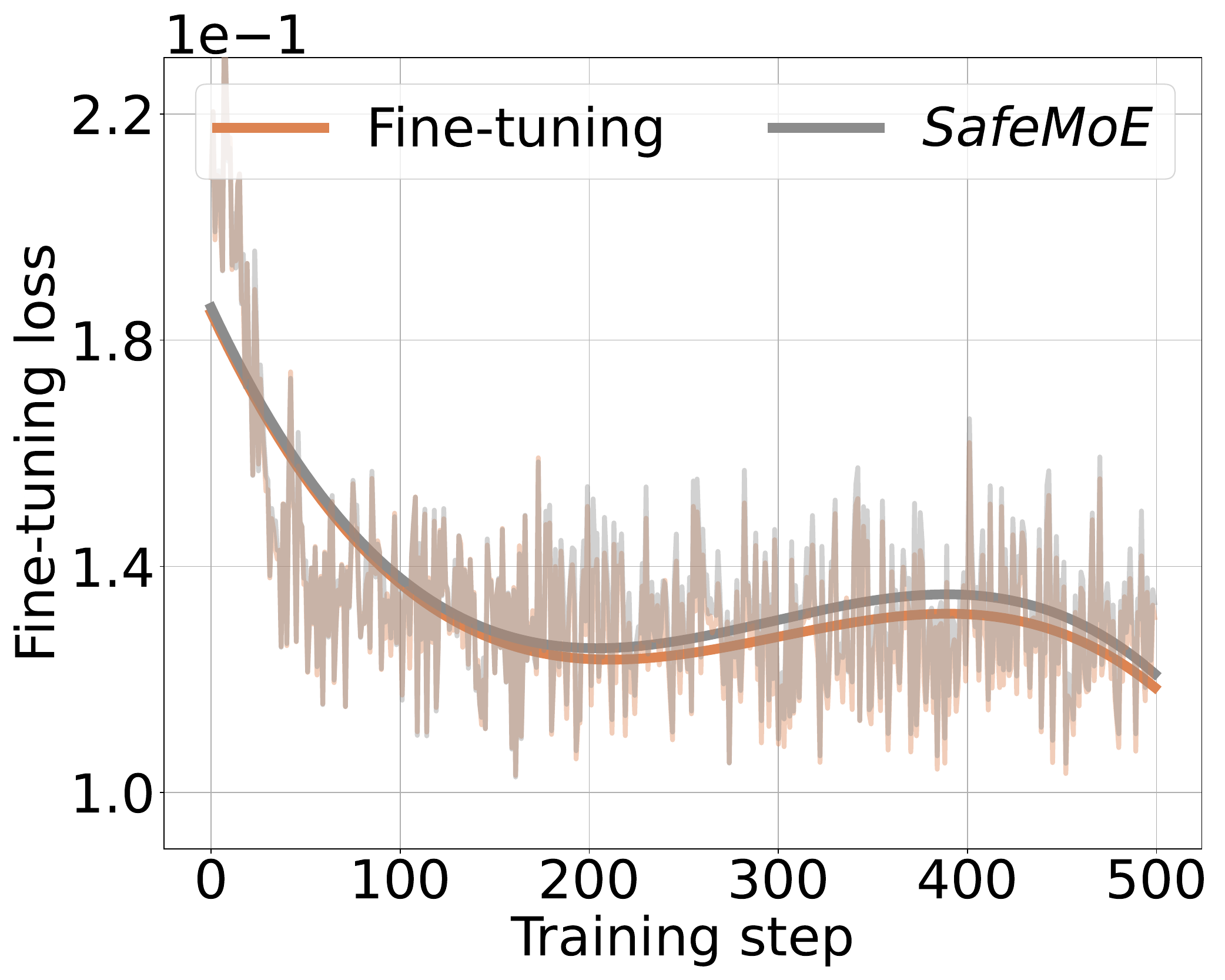}
        \caption{FT loss (w/ smoothing line)}
        \label{subfig:dynamics_loss_safemoe}
    \end{subfigure}
    \caption{Training dynamics of vanilla fine-tuning vs. \ours{} (OLMoE on SAMSum).}
    \label{fig:dynamics}
\end{figure}

\subsection{Training Dynamics}

To demonstrate that the design of \ours{} is valid in practice, we analyze the variation of the safety routing drift, harmfulness score, and fine-tuning loss throughout the fine-tuning steps. Figure~\ref{fig:dynamics} presents the results for OLMoE fine-tuned on the SAMSum task.

As shown in Figure~\ref{subfig:dynamics_drift_safemoe} and~\ref{subfig:dynamics_hs_safemoe}, \ours{} moderately reduces the safety routing drift at the early stage and effectively mitigates it after the first regularization period (after 150 steps). The harmfulness score decreases with drift mitigation, achieving a significant reduction compared to vanilla fine-tuning at subsequent checkpoints. Our experiment has shown that existing defenses have limited effectiveness in reducing the safety routing drift and harmfulness (see Figure~\ref{fig:drift_baseline}). \ours{} methodologically addresses this limitation and provides an effective defense for MoE LLMs.

In Figure~\ref{subfig:dynamics_loss_safemoe}, we also present the fine-tuning loss to further evaluate its impact on the fine-tuning task. The loss converges stably, closely matching the trajectory of vanilla fine-tuning with only negligible differences. This underscores the compatibility of \ours{} in standard fine-tuning, enabling safe and practical fine-tuning services.

\subsection{Efficiency Analysis}

% \begin{table}[t]
%     \centering
%     \caption{Execution overheads of safe fine-tuning methods.}
%     \small
%     \begin{tabular} {l c c c c c}
%         \toprule
%          & \textbf{SafeInstr} & \textbf{SaLoRA} & \textbf{Antidote} & \textbf{SafeDelta} & \textbf{\ours{}} \\\midrule
%         Training time overhead ($s$) & 171.15 & 747.56 & $5.67\pm0.41$ & $52.18\pm1.41$ & $17.26\pm0.07$ \\
%         \bottomrule 
%     \end{tabular}
%     \label{tab:overhead}
% \end{table}

\setlength{\intextsep}{0.5cm}%
\begin{wraptable}{r}{0.475\columnwidth}
\vspace{-0.5cm}
% \begin{table}[t]
    \centering
    \caption{Execution overheads (OLMoE).}
    \small
    \begin{tabular} {l c c}
        \toprule
        \multirow{2.5}{*}{\textbf{Method}} & \multicolumn{2}{c}{\textbf{Extra time}} \\\cmidrule{2-3}
        & \textbf{Seconds} & \textbf{Percentage} \\ \midrule
        SafeInstr & 15.98 &1.98\% \\
        SaLoRA & 747.56 &92.41\% \\
        Antidote & 5.67 &0.70\% \\
        SafeDelta & 52.18 &6.45\% \\ \midrule
        \ours{} & 17.26 &2.13\% \\
        \bottomrule 
    \end{tabular}
    \label{tab:overhead}
% \end{table}
\end{wraptable}
\setlength{\intextsep}{12pt}

To assess the efficiency of \ours{}, we analyze the total execution overhead of \ours{} during fine-tuning and post-fine-tuning. Table~\ref{tab:overhead} summarizes results for OLMoE fine-tuned on the SAMSum task under the environment described in Appendix~\ref{appendix:exprimental_setting}. For reference, vanilla fine-tuning takes 808.98 seconds on average over ten trials. SaLoRA incurs substantial overhead in computing the optimal LoRA initialization, increasing the execution time to nearly twice that of vanilla fine-tuning. The other approaches are generally more efficient but fail to effectively mitigate HFT attacks on MoE LLMs. In contrast, \ours{} attains strong safety with only a 2.13\% increase in its training time, highlighting its practicality in safeguarding large-scale models.

\subsection{Layer-Selective Application of \ours{}}
\label{subsec:layer_selective}

\begin{figure}[t]
    \centering
    \begin{minipage}[t]{0.49\linewidth}
        \centering
        \includegraphics[width=\linewidth]{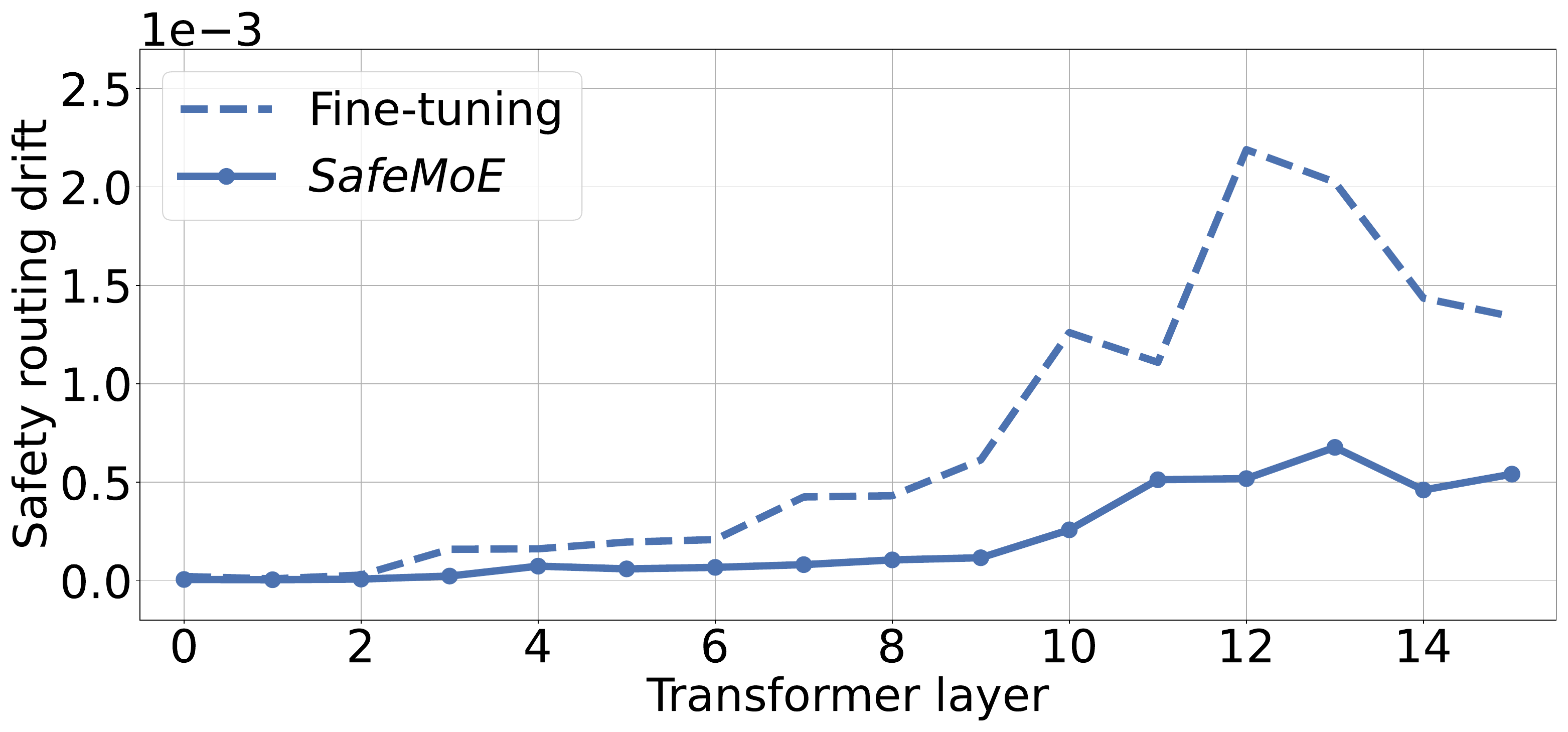}
        \caption{Safety routing drift across transformer layers (OLMoE on SAMSum).}
        \label{subfig:routing_drift_layer}
    \end{minipage}
    \hspace{0.1cm}
    \begin{minipage}[t]{0.49\linewidth}
        \centering
        \includegraphics[width=\linewidth]{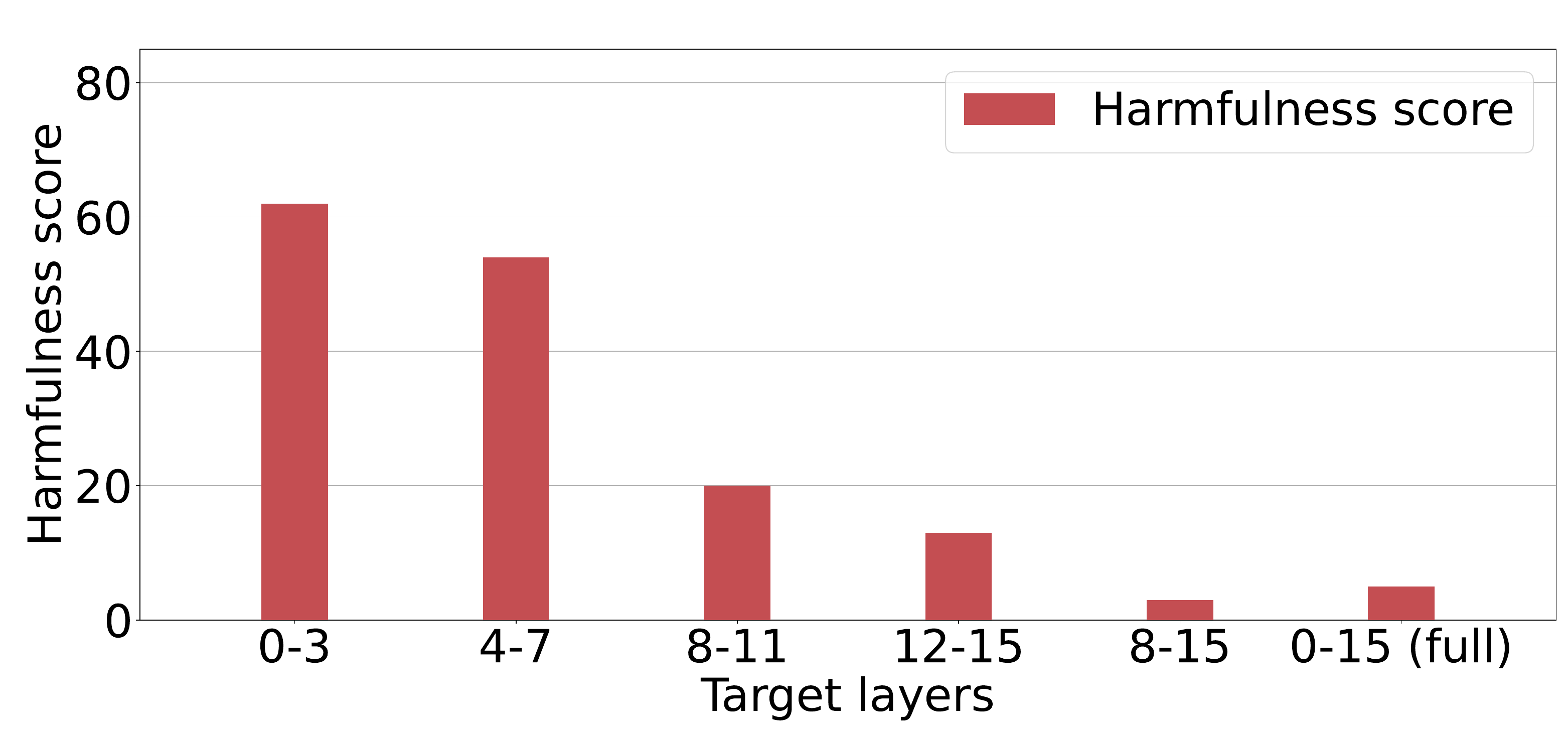}
        \caption{Effectiveness of \ours{} applied on specific layers (OLMoE on SAMSum).}
        \label{subfig:efficient_approach}
    \end{minipage}
\end{figure}

We analyze the extent of safety routing drift across transformer layers in OLMoE fine-tuned on the SAMSum task, as shown in Figure~\ref{subfig:routing_drift_layer}. The results reveal that routing drift is not uniform across layers and the upper layers exhibit substantially larger drift. This appears to result from harmful features typically being distinguished after the middle layers of LLMs~\citep{li2025safety}.

Motivated by this observation, we investigate more efficient variants of \ours{} that apply the safety routing drift regularization selectively rather than across all layers (Figure~\ref{subfig:efficient_approach}). The safety evaluation results show that targeting upper layers (e.g., 12-15 layers) provides much stronger mitigation of HFT attacks than targeting lower layers. Moreover, applying \ours{} only to 8-15 layers (the upper half) achieves a level of safety comparable to full-layer regularization.

\subsection{Sensitivity Analysis}

\begin{figure}[t]
    \centering
    \begin{subfigure}[t]{0.327\linewidth}
        \centering
        \includegraphics[width=\linewidth]{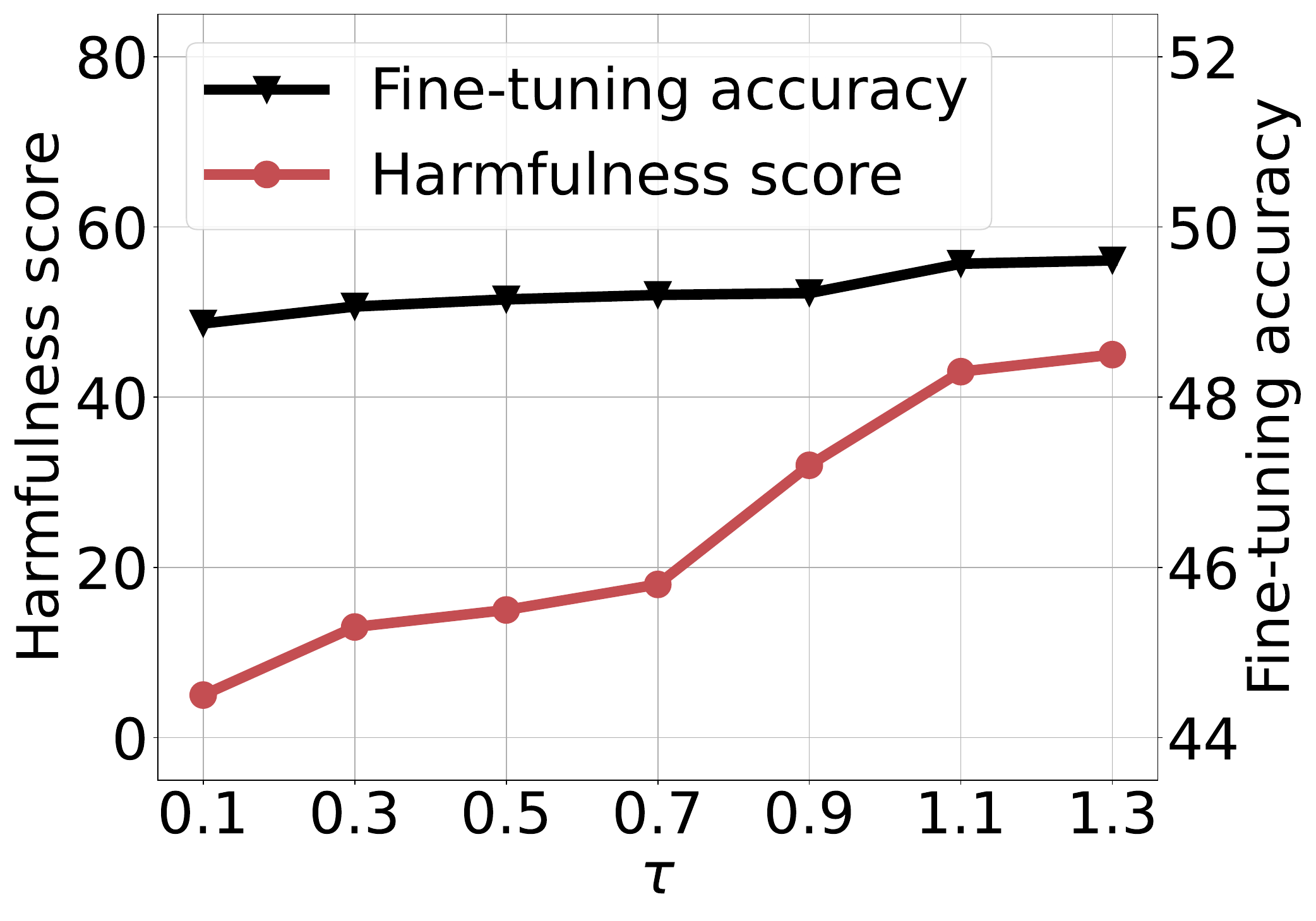}
        \caption{$\mathcal{L}_{\mathrm{reg}}$ hyperparameter ($\tau$)}
        \label{subfig:sensitivity_temp}
    \end{subfigure}
    % \hspace{0.2cm}
    \begin{subfigure}[t]{0.327\linewidth}
        \centering
        \includegraphics[width=\linewidth]{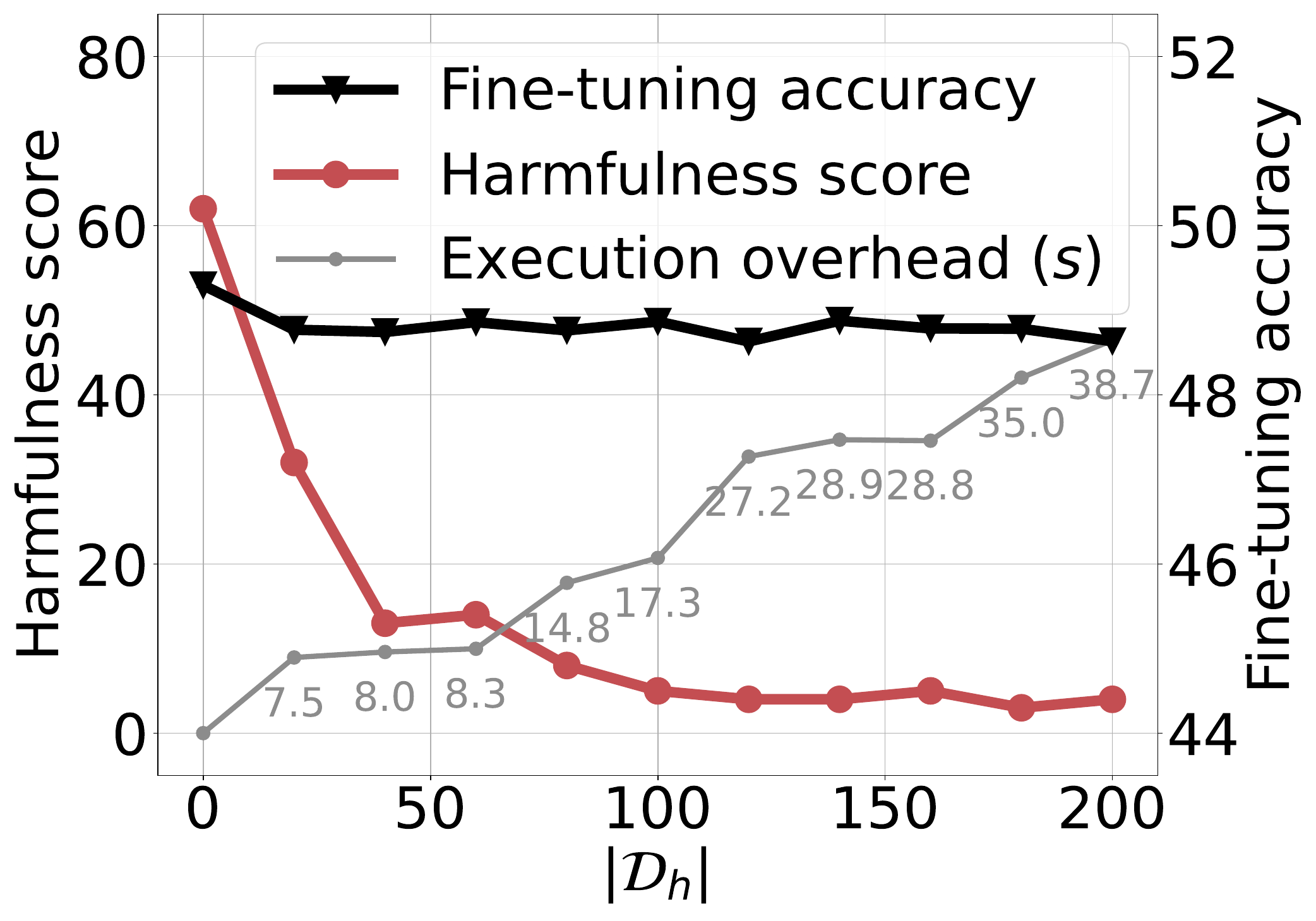}
        \caption{\# harmful instructions ($|\mathcal{D}_h|$)}
        \label{subfig:sensitivity_safe_size}
    \end{subfigure}
    % \hspace{0.2cm}
    \begin{subfigure}[t]{0.327\linewidth}
        \centering
        \includegraphics[width=\linewidth]{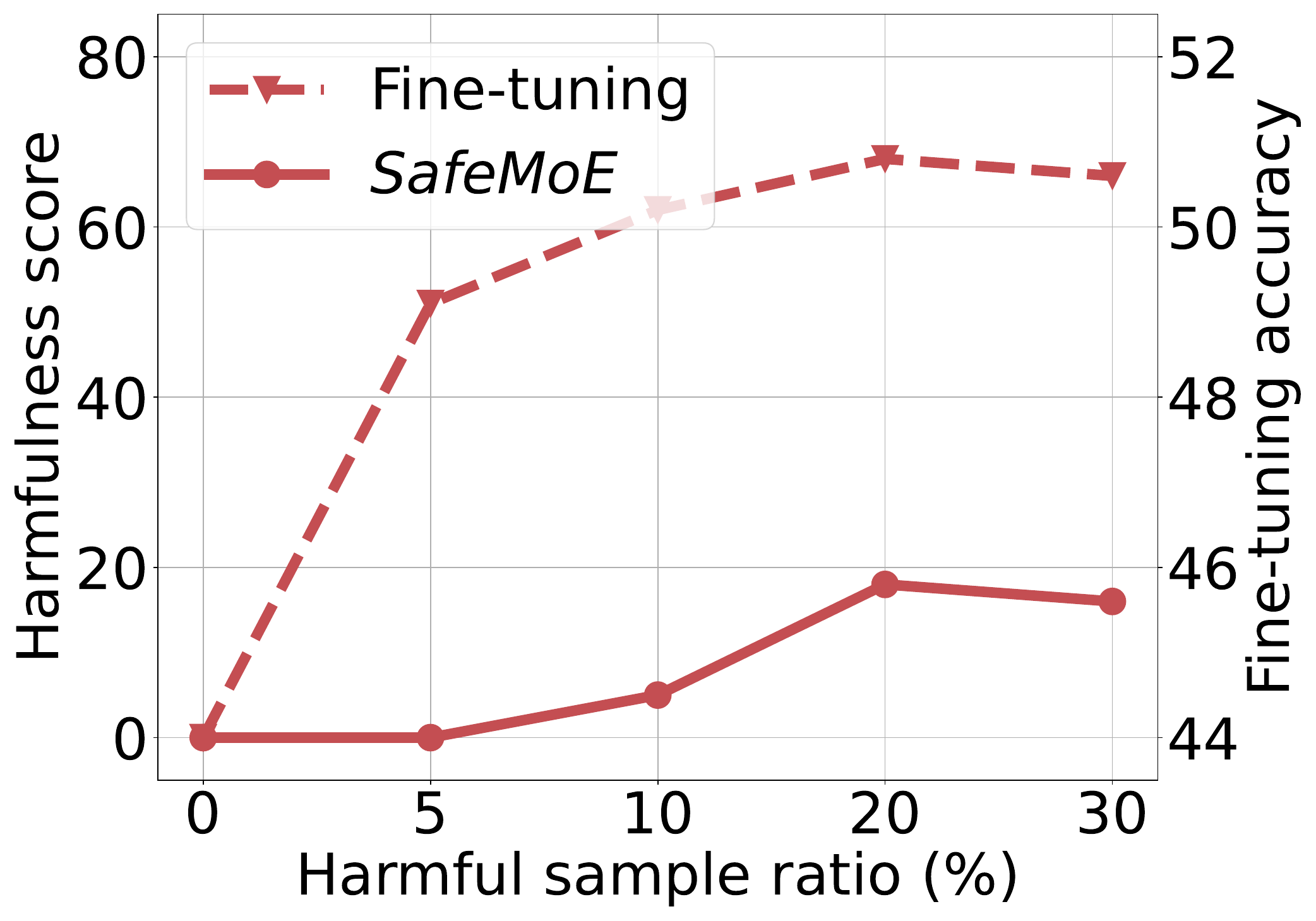}
        \caption{Harmful sample ratio in $\mathcal{D}_{ft}$}
        \label{subfig:sensitivity_harmful}
    \end{subfigure}
    \caption{Sensitivity to regularization strength (a, b) and attack strength (c) (OLMoE on SAMSum).}
    \label{fig:sensitivity}
    \vspace{-0.1cm}
\end{figure}

\noindent\textbf{Regularization hyperparameter.} The hyperparameter $\tau$ in Equation~\ref{eq:regularization} controls the strength of the safety routing drift regularization. Figure~\ref{subfig:sensitivity_temp} shows results for OLMoE fine-tuned with the SAMSum task. Smaller values of $\tau$ focus the regularization on top-ranked safety-critical experts, achieving substantial harmfulness reduction. Across $\tau$ values, fine-tuning accuracy remains nearly unchanged, demonstrating that \ours{} can robustly improve safety without sacrificing task utility.

\noindent\textbf{Number of harmful instructions.}
We vary the number of harmful instructions $|\mathcal{D}_h|$ used for the regularization (see Figure~\ref{subfig:sensitivity_safe_size}). More harmful instructions strengthen the safety effect with an approximately linear overhead increase, suggesting a simple yet effective way to enhance our approach. Our default choice ($|\mathcal{D}_h| = 100$) provides a practical balance between safety and efficiency.

\noindent\textbf{Harmful sample ratio in fine-tuning data.} We explore different strengths of HFT attacks by adjusting the ratio of harmful samples in the fine-tuning dataset $\mathcal{D}_{ft}$ (see Figure~\ref{subfig:sensitivity_harmful}). Vanilla fine-tuning exhibits drastically increasing harmfulness scores as the harmful ratio rises. In contrast, \ours{} suppresses harmfulness escalation, demonstrating robustness even against stronger attacks.

\subsection{Effectiveness under Full Fine-Tuning}

We extend our evaluation to a full-parameter fine-tuning scenario to demonstrate the generality of \ours{}.
%
% In this setting, all model layers, including self-attention, experts, and gating network layers, are trainable, while the LoRA-based main experiments train only the LoRA layers inserted to self-attention layers.
%
Table~\ref{tab:full_ft} shows the results for OLMoE fine-tuned on the SAMSum task. \ours{} achieves substantial harmfulness reduction while preserving fine-tuning accuracy in this setting as well. Although all expert layers, including safety-critical ones, are exposed to training during HFT attacks, \ours{} remains robustly effective by solely preventing the safety routing drift. This further highlights the central role of routing in the safety of MoE LLMs and confirms the validity of our approach. Moreover, it incurs only a 2.30\% increase in execution time (four GPUs), comparable to that observed in the LoRA-based setting (2.13\% with one GPU). These results show that \ours{} is reliably adaptable to both parameter-efficient and full fine-tuning approaches.

\subsection{Evaluation on Additional Harmfulness Benchmark}

To further validate our findings, we evaluate harmfulness of fine-tuned MoE LLMs on HEx-PHI~\citep{qi2024fine}, a widely used harmful instruction benchmark. Table~\ref{tab:hex_phi} reports results for the SAMSum fine-tuning scenario under our main settings. The results remain consistent, showing that \ours{} significantly outperforms all baselines, while SafeInstr achieves moderate harmfulness reduction with strong performance for Qwen1.5 MoE. These findings confirm the robust defense effectiveness of \ours{} against diverse harmful querying scenarios.

% \begin{table}[t]
%     \centering
%     \caption{Safety evaluation under full fine-tuning, shown in comparison with our LoRA-based results.}
%     \small
%     \begin{tabular} {l c c c c c}
%         \toprule
%         \multirow{3}{*}{Method} & \multicolumn{2}{c}{\textbf{LoRA (in Table~\ref{tab:eval_main})}} && \multicolumn{2}{c}{\textbf{Full Fine-Tuning}} \\ \cmidrule{2-3}\cmidrule{5-6}
%         & FA & HS && FA & HS \\ \midrule
%         Aligned & 31.8 & 0  && 31.8 & 0 \\
%         Fine-Tuning & 49.3 & 62.0 && \\ \midrule
%         \ours{} & 48.9 & 5.0 && 51.0 & 2.0 \\ \bottomrule
%     \end{tabular}
%     \label{tab:full_ft}
% \end{table}

% \begin{table}[t]
%     \centering
%     \caption{Safety evaluation under full fine-tuning, shown in comparison with our LoRA-based results.}
%     \small
%     \begin{tabular} {l c c c c c c c c}
%         \toprule
%         \multirow{2.5}{*}{Method} & \multicolumn{3}{c}{\textbf{LoRA (in Table~\ref{tab:eval_main})}} && \multicolumn{3}{c}{\textbf{Full fine-tuning}} \\ \cmidrule{2-4}\cmidrule{6-8}
%         & FA & HS & Training time ($s$) (1 GPU) && FA & HS & Training time ($s$) (4 GPUs) \\ \midrule
%         Aligned & 31.8 & 0 & - && 31.8 & 0 & -\\
%         Fine-tuning & 49.3 & 62.0 & 808.98 && 50.4 & 58.0 & 7,023.79 \\ \midrule
%         \ours{} & 48.9 & 5.0 & 826.25 (+17.26) && 51.0 & 2.0 &  7,189.01 (+165.21)  \\ \bottomrule
%     \end{tabular}
%     \label{tab:full_ft}
% \end{table}

% lora safemoe 826.25 (+17.26)
% full safemoe 7189.01 (+165.2142952) 

\begin{table}[t]
\centering
    \begin{minipage}[t]{0.4\linewidth}
        \centering
        \caption{Full fine-tuning results of fine-tuning accuracy (FA$\uparrow$), harmfulness score (HS$\downarrow$), and training time. }
        \vspace{-0.05cm}
        \small
        \begin{tabular} {l c c c}
            \toprule
            \multirow{3.5}{*}{\textbf{Method}} & \multicolumn{3}{c}{\textbf{OLMoE on SAMSum}} \\ \cmidrule{2-4}
            & \textbf{FA} & \textbf{HS} & \textbf{Time($s$)} \\ \midrule
            Aligned & 31.8 & 0 & -\\
            Fine-tuning & 50.4 & 58.0 & 7,023.79 \\ \midrule
            \ours{} & 51.0 & 2.0 &  \makecell{7,189.01\\(+2.30\%)}  \\ \bottomrule
        \end{tabular}
        \label{tab:full_ft}
    \end{minipage}
    \hspace{0.3cm}
    \begin{minipage}[t]{0.55\linewidth}
        \centering
        \caption{Harmfulness score (HS$\downarrow$) on HEx-PHI.}
        \small
        \begin{tabular} {l c @{\quad} c @{\quad} c}
            \toprule
            \textbf{Method} & \textbf{OLMoE} & \textbf{Qwen1.5 MoE} & \textbf{DeepSeek V2} \\\midrule
            Aligned & 0.3 & 8.7 & 5.7 \\
            Fine-tuning & 79.7 & 33.0 & 83.0 \\ \midrule
            SafeInstr & 52.0 & \underline{13.3} & \underline{31.0} \\
            SaLoRA & 39.3 & 34.7 & 64.7\\
            Antidote & 53.3 & 24.7 & 68.0 \\
            SafeDelta & \underline{21.7} & 28.7 & 41.0\\ \midrule
            \ours{} & \textbf{6.3} & \textbf{10.0} & \textbf{3.3}\\ \bottomrule
        \end{tabular}
        \label{tab:hex_phi}
    \end{minipage}
\end{table}
\section{Related Work}

We categorize recent defenses against HFT attacks into three groups based on their application stage.

\textbf{Alignment stage.} These methods aim to enhance the robustness of the model against subsequent HFT attacks. Vaccine~\citep{huang2024vaccine} trains models to resist perturbations that maximize alignment loss. RepNoise~\citep{rosati2024representation} and Booster~\citep{huangbooster} proactively remove harmful information by optimizing perturbations in model representations or weights. VAA~\citep{liangvulnerability} introduces a vulnerability-aware alignment method that balances training across vulnerable and invulnerable subsets of alignment data. However, such alignment-stage methods require carefully tuned hyperparameter for each downstream task~\cite{huangbooster}, which limits their practicality in fine-tuning services that must handle many unknown tasks.

\textbf{Fine-tuning stage.} A second line of work directly addresses safety degradation during fine-tuning. SafeInstr~\citep{bianchi2024safety} augments supervised fine-tuning datasets with safe samples. Lisa~\citep{huang2024lazy} introduces a proximal optimization method to mitigate convergence instability when jointly training alignment and task-specific data. SAFT~\citep{choi2024safety} and SEAL~\citep{shenseal} identify and filter harmful samples from fine-tuning data by scoring their safety impact. SaLoRA~\citep{li2025salora} proposes a safety-aware initialization of LoRA layers, designed based on an analysis of changes in safety-related features observed during fine-tuning.

\textbf{Post-fine-tuning stage.} Training-free remedies have also been proposed to restore safety after harmful fine-tuning. RESTA~\citep{bhardwaj2024language} extracts a safety vector from an aligned model and reintroduces it into the fine-tuned model via arithmetic addition of weights. SafeLoRA~\citep{hsu2024safe} selectively projects LoRA weights into a safety-aligned subspace. Antidote~\citep{huang2025antidote} removes harmful parameters identified through their importance to alignment data, thereby improving robustness under varying fine-tuning hyperparameters. SafeDelta~\citep{lu2025safe} refines fine-tuned delta parameters to balance task utility with reduced safety degradation.

% Despite extensive research on safeguarding LLM fine-tuning, no prior work has systematically analyzed the safety vulnerabilities of MoE LLMs under HFT attacks or proposed tailored defenses.
\section{Conclusion}

This work introduces \ours{}, the first safe fine-tuning method tailored to MoE-based LLMs. Our systematic analysis uncovers a vulnerability inherent in their safety mechanisms; routing decisions for harmful inputs drift significantly from those of safety-aligned models under both harmful and benign fine-tuning. To address this, we propose a routing drift regularization method with an efficient optimization algorithm that integrates seamlessly into standard MoE LLM fine-tuning pipelines. Extensive evaluations across diverse MoE LLMs show that \ours{} achieves significant reductions in harmfulness with minimal overhead. These results establish \ours{} as an effective and practical defense for fine-tuning services against HFT attacks, underscoring the need to address architectural weakness when safeguarding MoE LLMs from fine-tuning risks.

% \clearpage
% \input{sections/supplementary}
% \clearpage

\bibliography{iclr2026_conference}
\bibliographystyle{iclr2026_conference}

\clearpage

\appendix
\section{Appendix}

\subsection{Bi-Level Greedy Optimization of \ours{}}

\begin{algorithm}[ht]
\renewcommand{\algorithmicrequire}{\textbf{Input:}}
\renewcommand{\algorithmicensure}{\textbf{Output:}}
\caption{Greedy optimization of safety routing drift regularization}
\label{alg:greedy}
\begin{algorithmic}[1]
\Require Safety-aligned MoE LLM $\vw_{align}$; Fine-tuning datasets $\mathcal{D}_{ft}$; Harmful instruction dataset $\mathcal{D}_{h}$; Total training steps $T$; Regularization period $T_{reg}$; Optimizer $\mathrm{Adam}(\eta,\beta_1,\beta_2,\epsilon)$
\Ensure The fine-tuned MoE LLM
\State Initialize model weights $\vw_{0} \gets \vw_{align}$
\State Precompute routing weights $\vr(x|\vw_{align})\, \forall x \in \mathcal{D}_{h}$
    \For{step $t \in T$}
        \State $\vg_{t} \gets \nabla_{\vw}\mathcal{L}_{sft}(\vw_{t})$ on $\mathcal{D}_{ft}$ 
        \State $\tilde\vw_{t+1} \gets \mathrm{Adam} (\vw_{t}, \vg_{t})$
        \If{$t\bmod T_{reg} = 0$}  \Comment{Run regularization every $T_{\text{reg}}$ steps}
            \For{batch $\mathcal{B}_{h} \subset \mathcal{D}_{h}$} 
                \State $\tilde\vg_{h} \gets \nabla_{\vw}\mathcal{L}_{reg}(\tilde\vw_{t+1})$, where $x \in \mathcal{B}_{h}$
                \State $\tilde\vw_{t+1} \gets \mathrm{Adam} (\tilde\vw_{t+1}, \tilde\vg_{h})$
            \EndFor
        \EndIf
        \State $\vw_{t+1} \gets \tilde\vw_{t+1}$
    \EndFor
% \Return $\vw_{T}$
\end{algorithmic}
\end{algorithm}

\subsection{Experimental Setting Details}
\label{appendix:exprimental_setting}

\noindent\textbf{System settings.} Our experiments were conducted in a GPU cloud instance equipped with 6 cores AMD EPYC 7H12, 192GB of RAM, and 1 to 4 NVIDIA A100 80GB GPUs, depending on the requirements of each experiment. For gpt-oss~\citep{openai2025gptoss}, we employed 4 NVIDIA H100 80GB GPUs due to its GPU architecture compatibility.

\noindent\textbf{Model specifications.} We summarize the specifications of MoE LLMs used in our experiments in Table~\ref{tab:models}.

\noindent\textbf{Fine-tuning details.} Fine-tuning is performed with LoRA~\citep{hu2022lora} using configurations detailed in Table~\ref{tab:lora_config}. We train for three epochs with a learning rate of $1e{-4}$ and a batch size of 32.

\noindent\textbf{Generation and prompt settings.} We use greedy decoding for all generations. For harmfulness evaluation, we adopt each model’s default system prompt if available, or \textit{``You are a helpful AI assistant.''} otherwise, with a summarized default system prompt for Llama 4. For zero-shot task utility evaluation, we use a customized task-specific system and user prompts. For the MMLU-Redux-2.0 task, we follow the user prompt in the Llama 3.1 evaluation ~\citep{llamaevals}. The system and user prompts used in our evaluation are shown in Table~\ref{tab:prompt_harmful} and Table~\ref{tab:prompt_task}.

\begin{table}[h]
    \centering
    \caption{Specifications of MoE LLMs used in our experiments.}
    \small
    \begin{tabular} {l c c c c}
        \toprule 
        \multirow{2}{*}{\textbf{Model}} & \textbf{\# layers} & \textbf{\# experts} & \multirow{2}{*}{\textbf{Top-$k$}} & \textbf{Parameters} \\
        & \textbf{(MoE + dense)} & \textbf{(routed + shared)} & & \textbf{(active / total)} \\ \midrule
        OLMoE-1B-7B-Instruct & 16 & 64 & 8 & 1.3B / 6.9B \\
        Qwen1.5-MoE-A2.7B-Chat & 24 & 60 + 4 & 4 & 2.7B / 14.3B \\
        DeepSeek-V2-Lite-Chat & 26 + 1 & 64 + 2 & 6 & 2.4B / 15.7B \\
        gpt-oss-20b & 24 & 32 & 4 & 3.6B / 20.9B \\
        Qwen3-30B-A3B & 48 & 128 & 8 & 3.3B / 30.5B \\
        Phi-3.5-MoE-instruct & 32 & 16 & 2 & 6.6B / 41.9B \\
        Llama-4-Scout-17B-16E-Instruct & 48 & 16 + 1 & 1 & 17B / 109B \\
        Mixtral-8x22B-Instruct-v0.1 & 56 & 22 & 2 & 39B / 141B \\ \bottomrule
    \end{tabular}
    \label{tab:models}
\end{table}

\begin{table}[ht]
    \centering
    \caption{LoRA configurations for fine-tuning.}
    \small
    \begin{tabular} {l c c c c}
        \toprule 
        \textbf{Model} & \textbf{Target modules} & \textbf{Rank ($r$)} & \textbf{Alpha ($\alpha$)} & \textbf{Trainable parameters} \\\midrule
        OLMoE-1B-7B-Instruct & q, v & 8 & 8 & 1.0M (0.0152\%)\\
        Qwen1.5-MoE-A2.7B-Chat & q, k, v, o & 8 & 32 & 3.1M (0.0220\%) \\
        DeepSeek-V2-Lite-Chat & q, kv\_a, kv\_b, o & 8 & 32 & 3.6M (0.0226\%) \\
        gpt-oss-20b & q, k, v, o & 8 & 16 & 4.0M (0.0190\%)\\
        Qwen3-30B-A3B & q, k, v, o & 8 & 32 & 3.3M (0.0109\%)\\
        Phi-3.5-MoE-instruct & q, k, v, o & 8 & 32 & 6.8M (0.0163\%)\\
        Llama-4-Scout-17B-16E-Instruct & q, k, v, o & 8 & 32 & 12.6M (0.0116\%) \\
        Mixtral-8x22B-Instruct-v0.1 & q, k, v, o & 8 & 32 & 17.4M (0.0124\%) \\ \bottomrule
    \end{tabular}
    \label{tab:lora_config}
\end{table}

\subsection{Baseline Tuning}
\label{appendix:baseline_tuning}

We extensively tune the hyperparameters of baseline methods for safeguarding MoE-based LLMs. The results of OLMoE on the SAMSum and SQL tasks are shown in Table~\ref{tab:baseline_tuning_samsum} and Table~\ref{tab:baseline_tuning_sql}, respectively. For each baseline, we select the hyperparameter setting that exhibits the lowest harmfulness score while allowing up to a 1\% degradation in fine-tuning accuracy.

\begin{table}[ht!]
    \centering
    \caption{Baseline tuning results of OLMoE on SAMSum. The selected ones are underlined.}
    \small
    \begin{tabular} {l @{} c c c c @{} c c c c @{} c c c c}
        \toprule
        & \textbf{Fine-tuning} & \multicolumn{3}{c}{\textbf{SaLoRA ($r_s=r_t$)}} && \multicolumn{3}{c}{\textbf{Antidote ($\alpha$)}} && \multicolumn{3}{c}{\textbf{SafeDelta ($s$)}} \\ \cmidrule{3-5} \cmidrule{7-9} \cmidrule{11-13}
        Hyperparameters & - & \underline{64} & 32 & 16 && 0.02 & \underline{0.03} & 0.04 && 2900 & \underline{2800} & 2700 \\\midrule
        Harmfulness score & 62.0 & 24.0 & 24.0 & 17.0 && 45.0 & 40.0 & 18.0 && 18.0 & 13.0 & 12.0 \\
        Fine-tuning accuracy & 49.3 & 48.9 & 48.3 & 48.1 && 49.3 & 48.7 & 48.1 && 49.0 & 48.6 & 47.5 \\
        \bottomrule 
    \end{tabular}
    \label{tab:baseline_tuning_samsum}
\end{table}
% \vspace{-0.2cm}
\begin{table}[ht!]
    \centering
    \caption{Baseline tuning results of OLMoE on SQL. The selected ones are underlined.}
    \small
    \begin{tabular} {l @{} c c c c @{} c c c c @{} c c c c}
        \toprule
        & \textbf{Fine-tuning} & \multicolumn{3}{c}{\textbf{SaLoRA ($r_s=r_t$)}} && \multicolumn{3}{c}{\textbf{Antidote ($\alpha$)}} && \multicolumn{3}{c}{\textbf{SafeDelta ($s$)}} \\ \cmidrule{3-5} \cmidrule{7-9} \cmidrule{11-13}
        Hyperparameters & - & 64 & 32 & \underline{16} && \underline{0.01} & 0.02 & 0.03 && 2900 & 2800 & \underline{2700} \\\midrule
        Harmfulness score & 64.0 & 48.0 & 40.0 & 25.0 && 44.0 & 36.0 & 40.0 && 52.0 & 38.0 & 33.0 \\
        Fine-tuning accuracy & 58.5 & 53.6 & 54.5 & 53.6 && 57.5 & 56.8 & 56.7 && 57.4 & 57.4 & 57.4 \\
        \bottomrule 
    \end{tabular}
    \label{tab:baseline_tuning_sql}
\end{table}

\subsection{Activation Probability of Safety-Critical Experts}
\label{appendix:activation}

\begin{figure}[ht]
    \centering
    \begin{subfigure}[t]{0.327\linewidth}
        \centering
        \includegraphics[width=\linewidth]{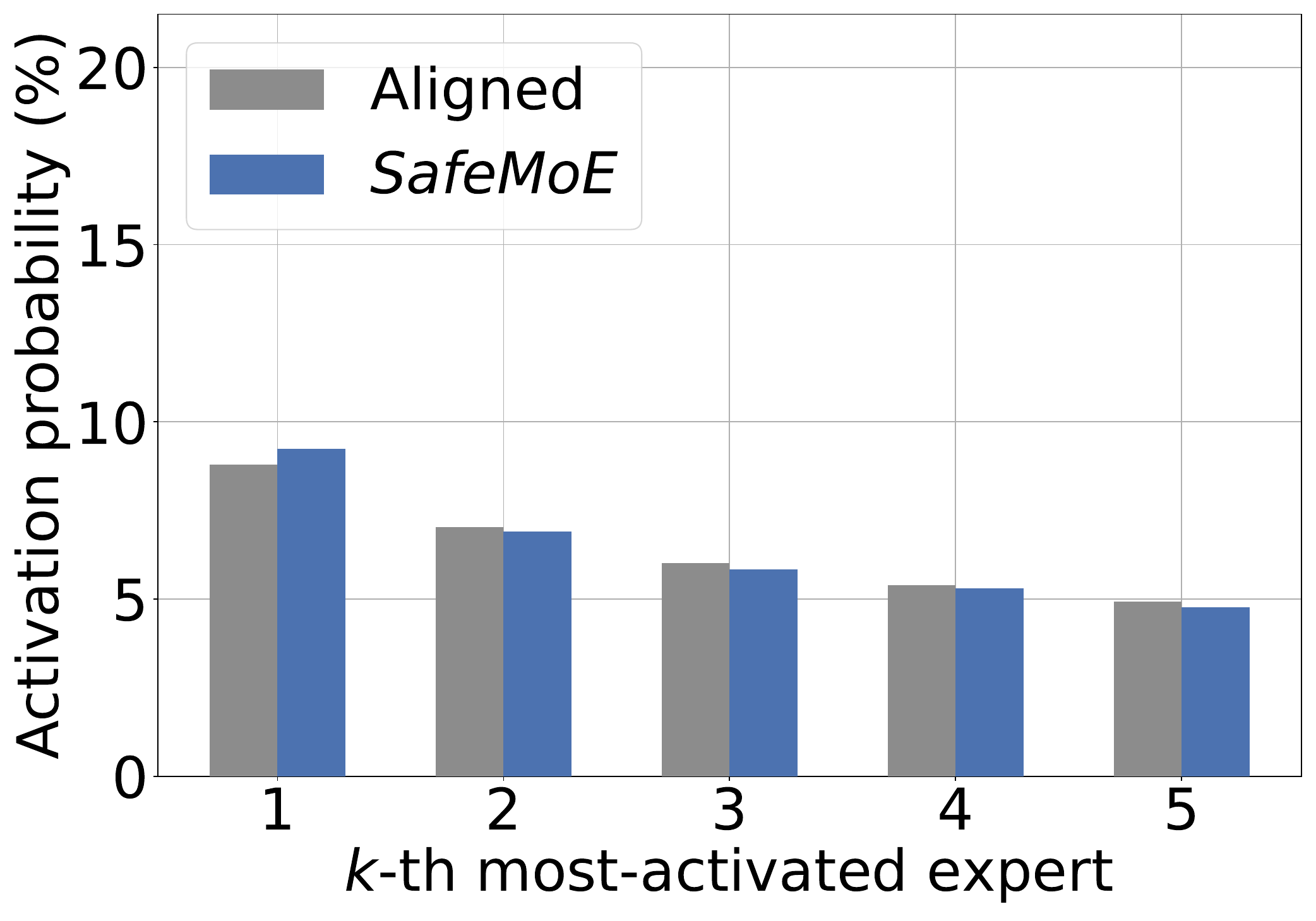}
        \caption{OLMoE (SAMSum)}
    \end{subfigure}
    % \hspace{0.2cm}
    \begin{subfigure}[t]{0.327\linewidth}
        \centering
        \includegraphics[width=\linewidth]{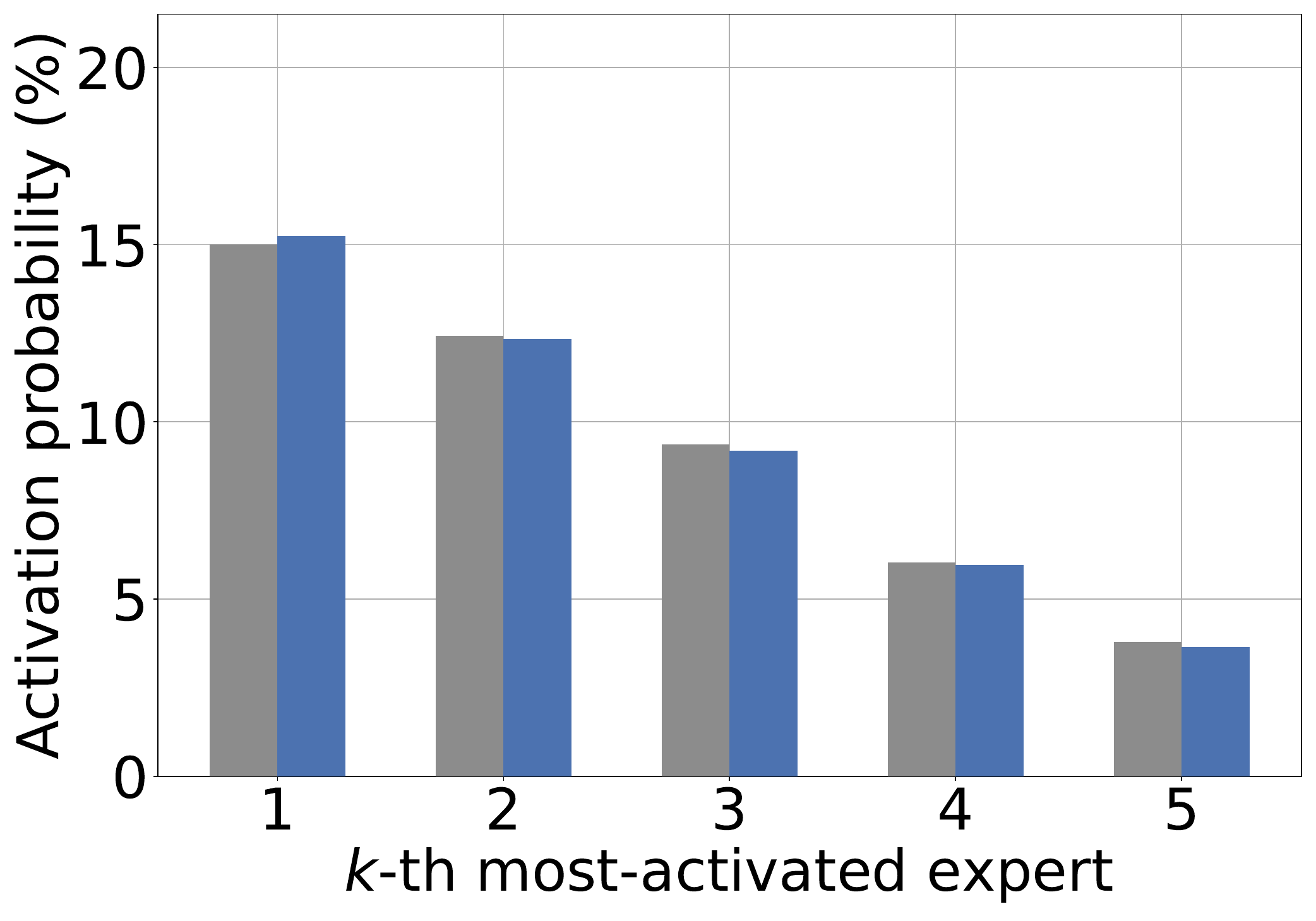}
        \caption{Qwen1.5 MoE (SAMSum)}
    \end{subfigure}
    % \hspace{0.2cm}
    \begin{subfigure}[t]{0.327\linewidth}
        \centering
        \includegraphics[width=\linewidth]{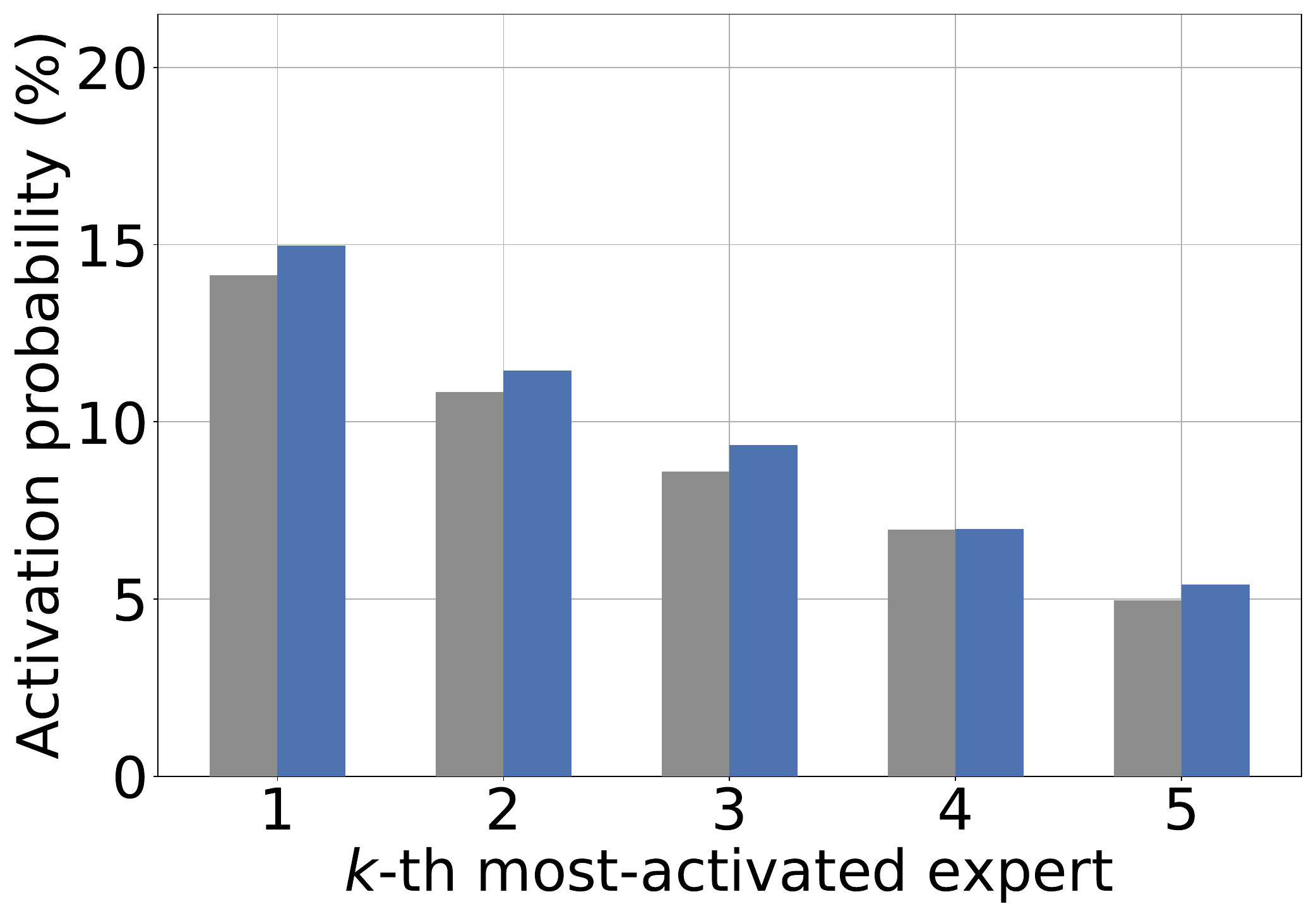}
        \caption{DeepSeek V2 (SAMSum)}
    \end{subfigure}
    \begin{subfigure}[t]{0.327\linewidth}
        \centering
        \includegraphics[width=\linewidth]{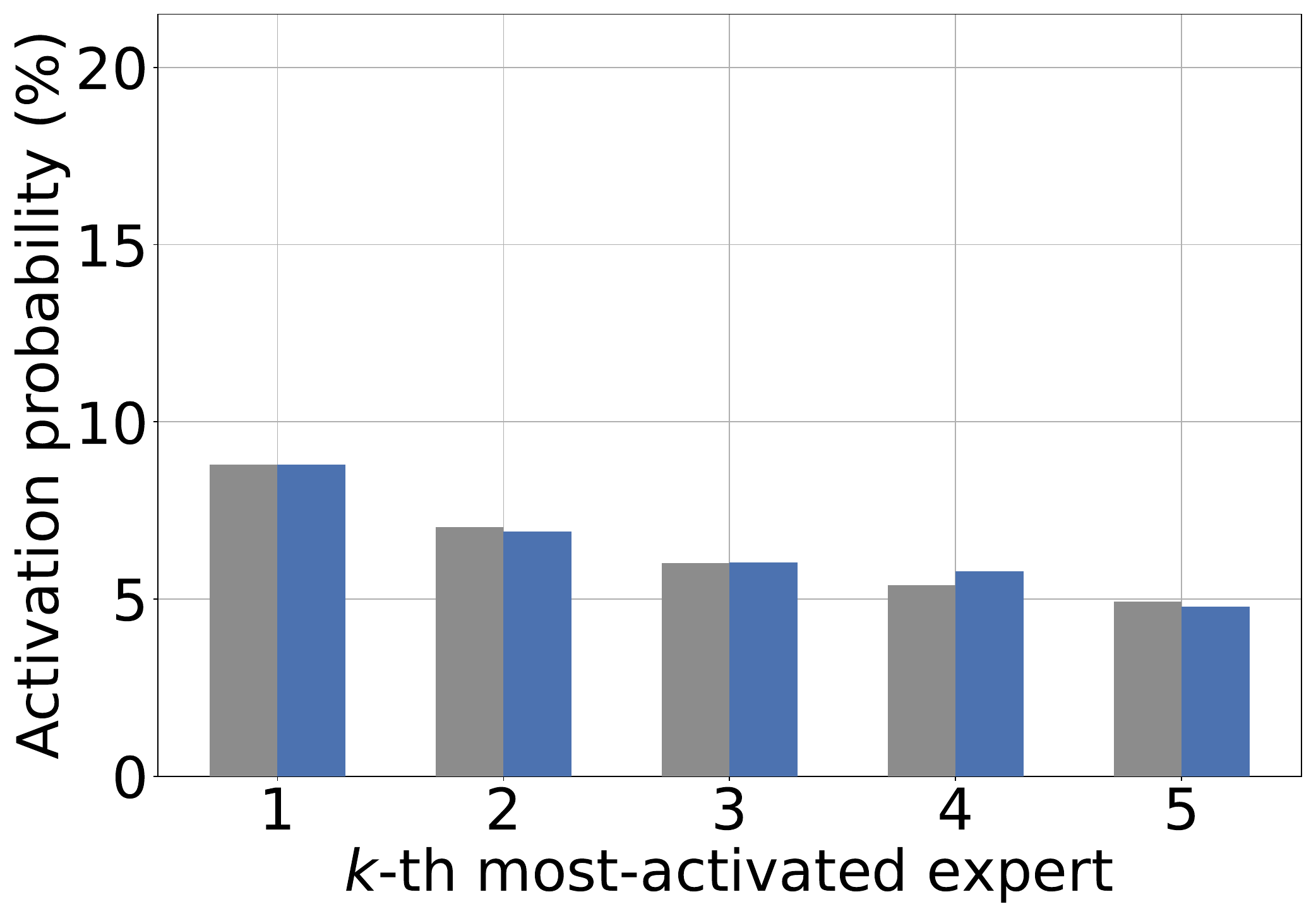}
        \caption{OLMoE (SQL)}
    \end{subfigure}
    % \hspace{0.2cm}
    \begin{subfigure}[t]{0.327\linewidth}
        \centering
        \includegraphics[width=\linewidth]{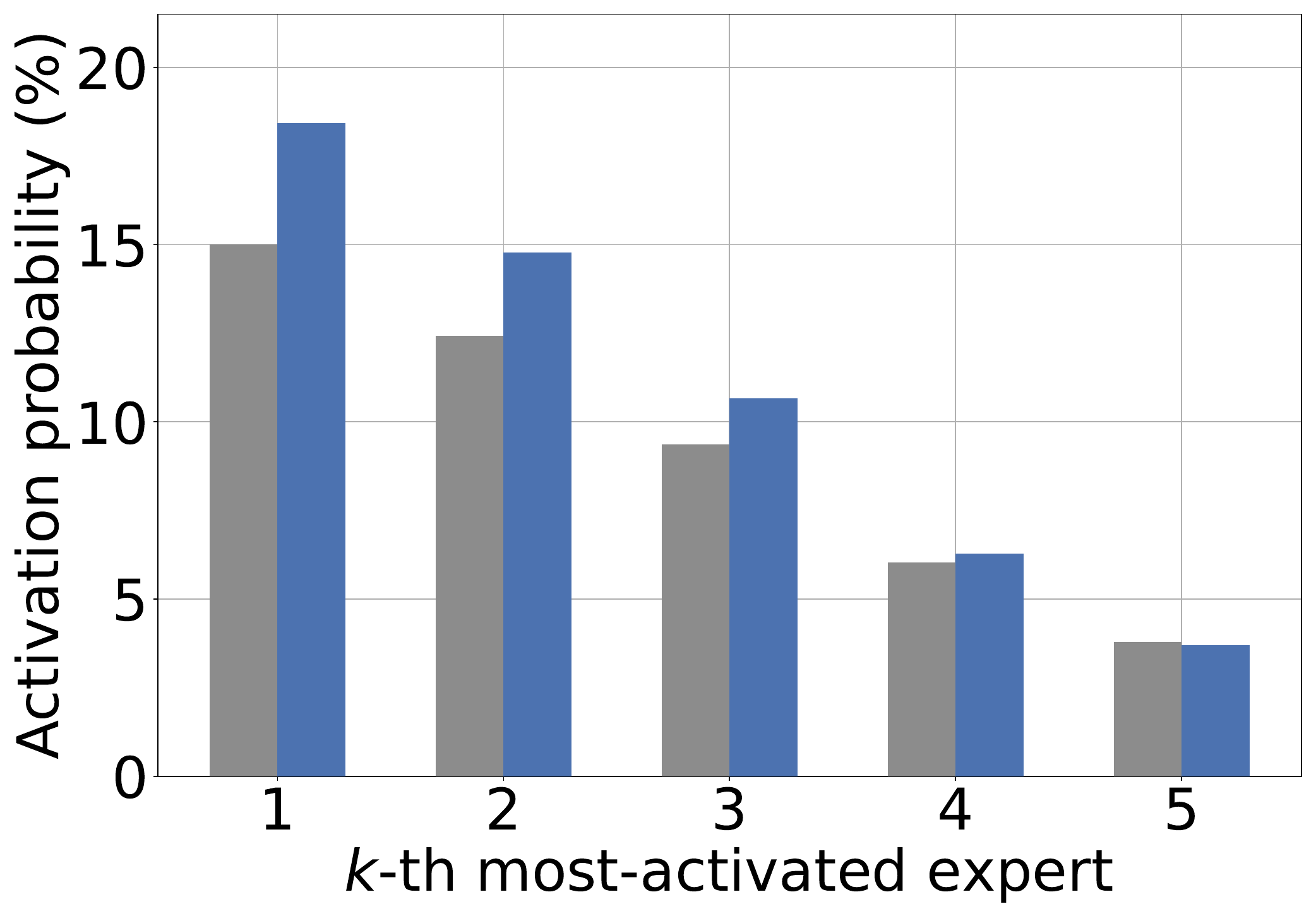}
        \caption{Qwen1.5 MoE (SQL)}
    \end{subfigure}
    % \hspace{0.2cm}
    \begin{subfigure}[t]{0.327\linewidth}
        \centering
        \includegraphics[width=\linewidth]{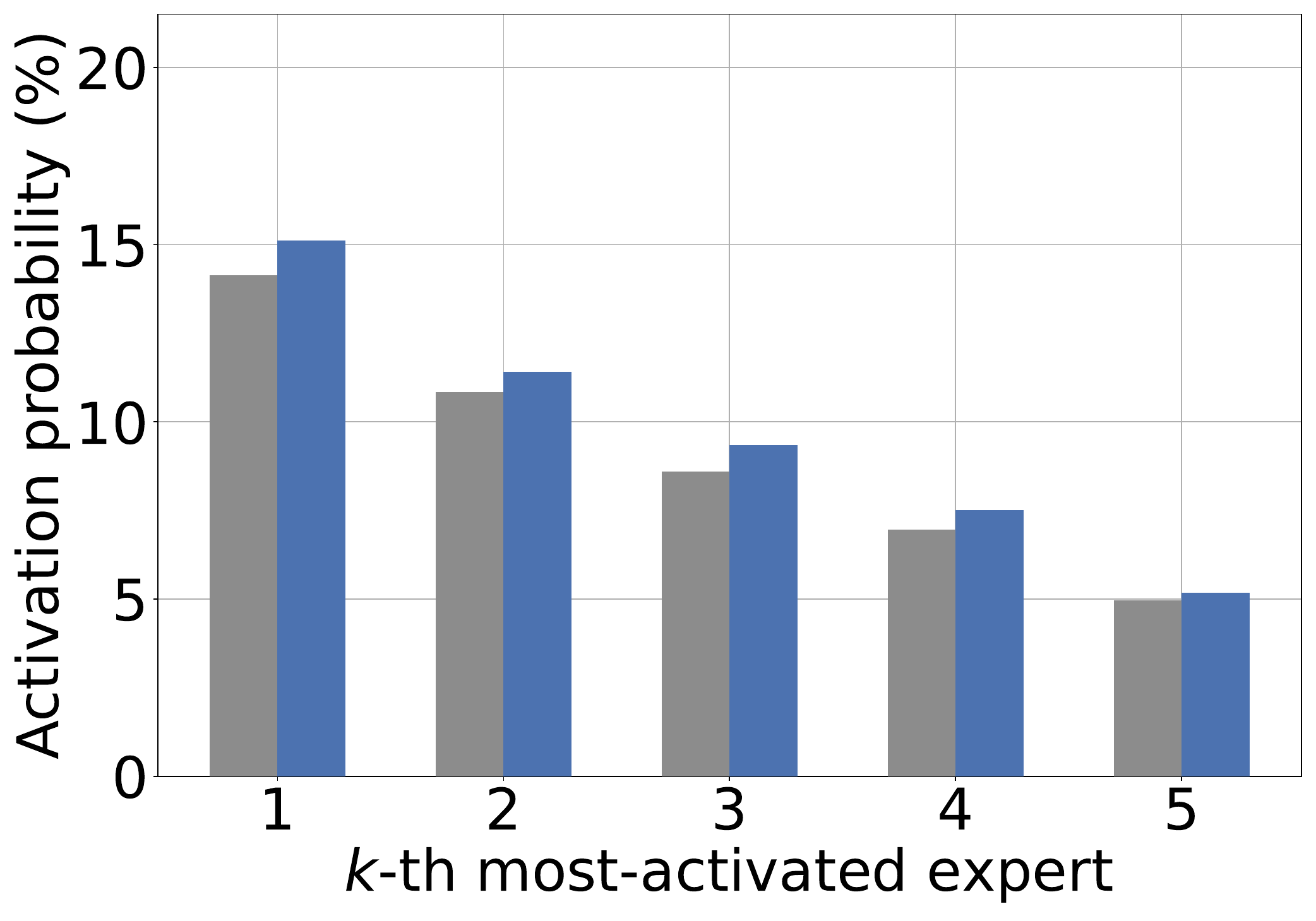}
        \caption{DeepSeek V2 (SQL)}
    \end{subfigure}
    \caption{Activation probability of top-ranked experts for harmful instructions, ranked by their probabilities in the safety-aligned models.}
    \label{fig:activation}
\end{figure}

We analyze the activation probabilities of experts when processing harmful instructions. These probabilities are obtained by applying Softmax to the routing weights. The top-ranked experts serve as safety-critical experts. Figure~\ref{fig:activation} compares their activation probabilities in the initial safety-aligned models and in the fine-tuned models with \ours{}. We find that \ours{} further increases the activation of safety-critical experts in the fine-tuned models. One possible explanation is that although \ours{} aims to resemble the routing decisions of the safety-aligned model, it learns to assign larger routing weights to safety-critical experts rather than simply replicating their original values. This can lead to slight improvements in safety compared to the initial safety-aligned models, as observed in our safety evaluation results in Table~\ref{tab:eval_main} and Table~\ref{tab:eval_large}.

\subsection{Layer-Wise Analysis of Routing Drift across Baselines}
\label{appendix:drift_layer}

\begin{figure}[ht]
    \centering
    \includegraphics[width=0.9\linewidth]{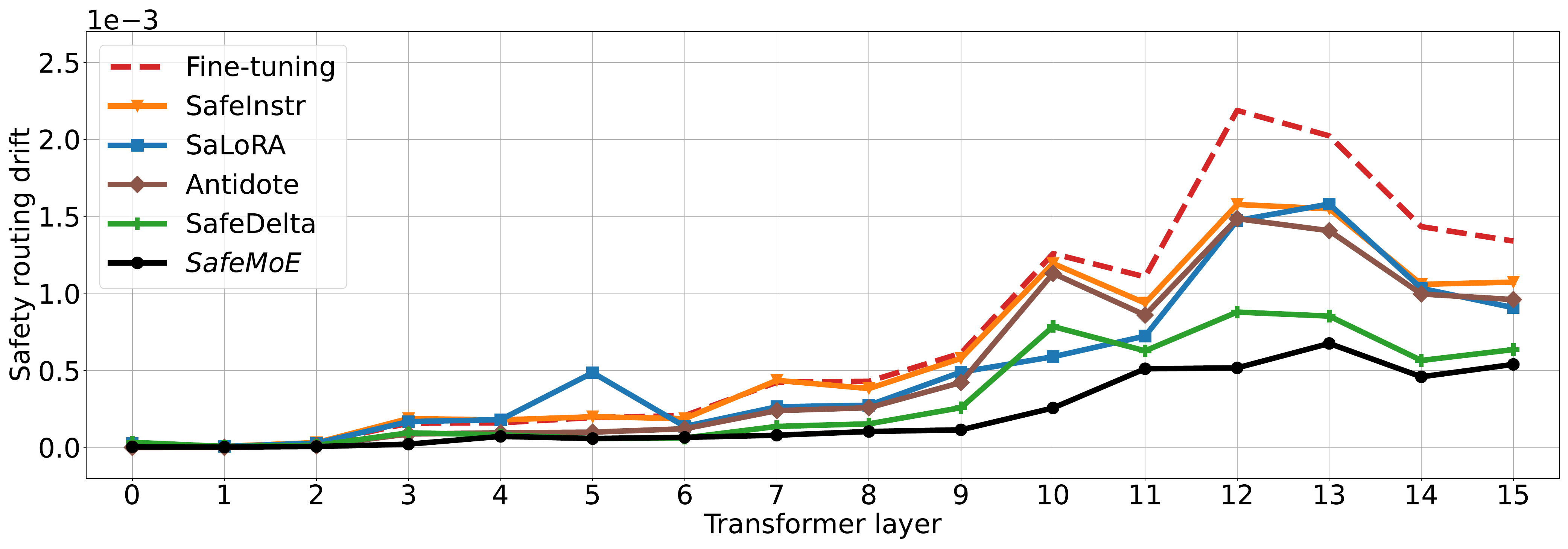}
    \caption{Safety routing drift of fine-tuned models across baseline methods (OLMoE on SAMSum).}
    \label{fig:drift_baseline}
\end{figure}

We compare the safety routing drift across transformer layers under the baseline methods. Figure~\ref{fig:drift_baseline} shows the results of OLMoE fine-tuned on the SAMSum task. The baselines consistently fail to address the substantial drift significant in the upper layers. In contrast, \ours{} directly mitigates it, thereby safeguarding MoE LLMs against HFT attacks. These results highlight the importance of an architecture-aware design and demonstrate the effectiveness of \ours{} in ensuring safety.

\subsection{Safety Category}

\begin{figure}[ht]
    \centering
    \begin{subfigure}[t]{0.327\linewidth}
        \centering
        \includegraphics[width=\linewidth]{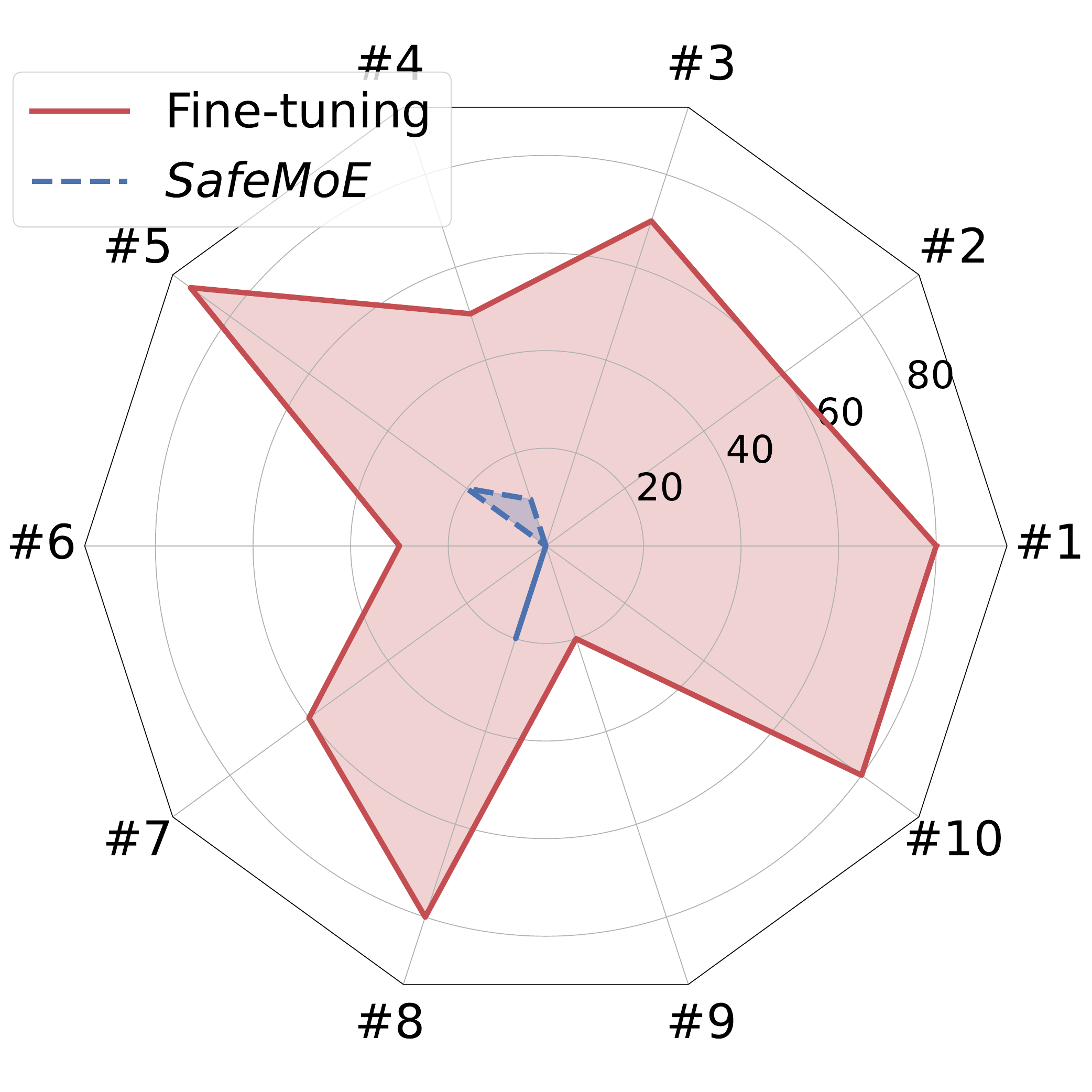}
        \caption{OLMoE (SAMSum)}
    \end{subfigure}
    % \hspace{0.2cm}
    \begin{subfigure}[t]{0.327\linewidth}
        \centering
        \includegraphics[width=\linewidth]{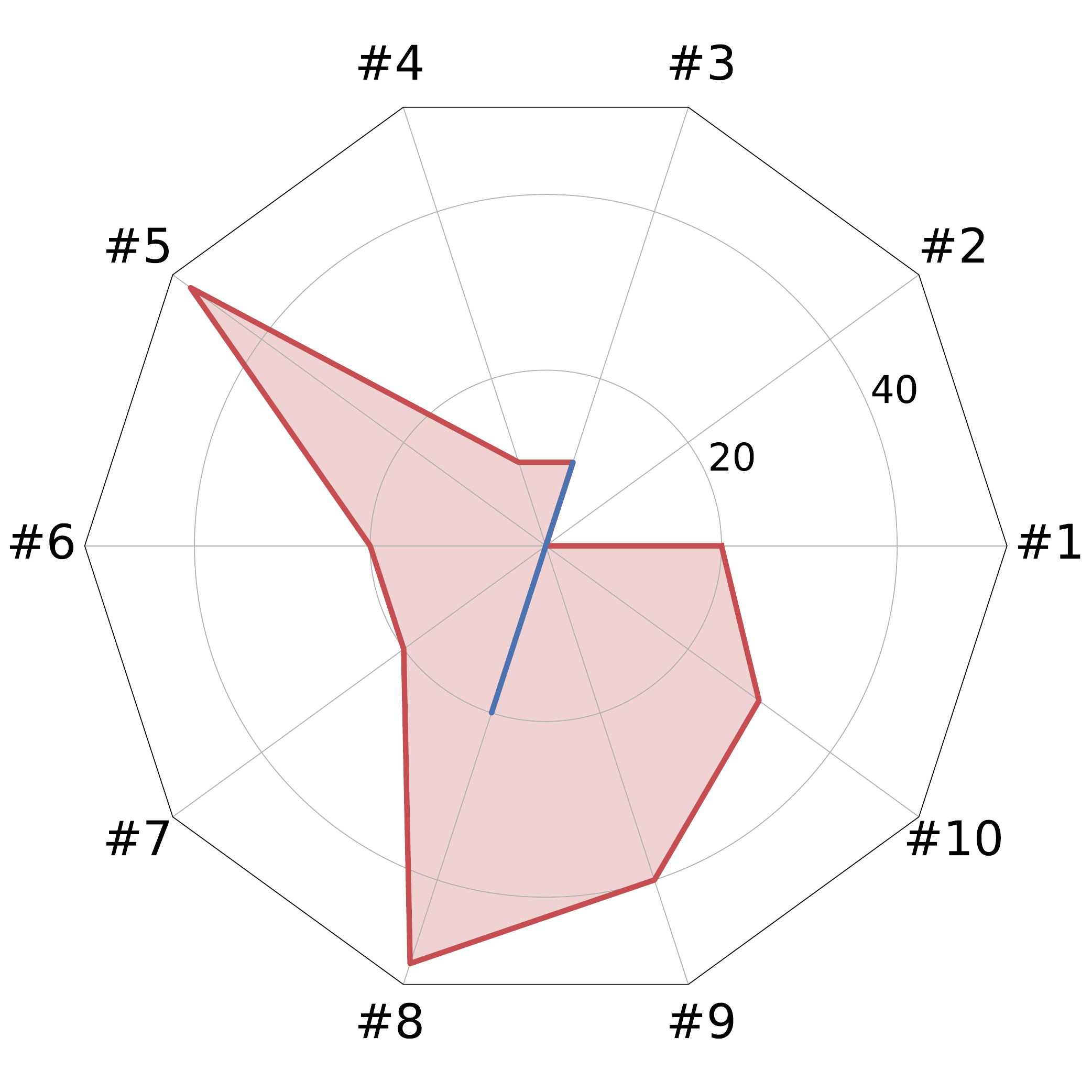}
        \caption{Qwen1.5 MoE (SAMSum)}
    \end{subfigure}
    % \hspace{0.2cm}
    \begin{subfigure}[t]{0.327\linewidth}
        \centering
        \includegraphics[width=\linewidth]{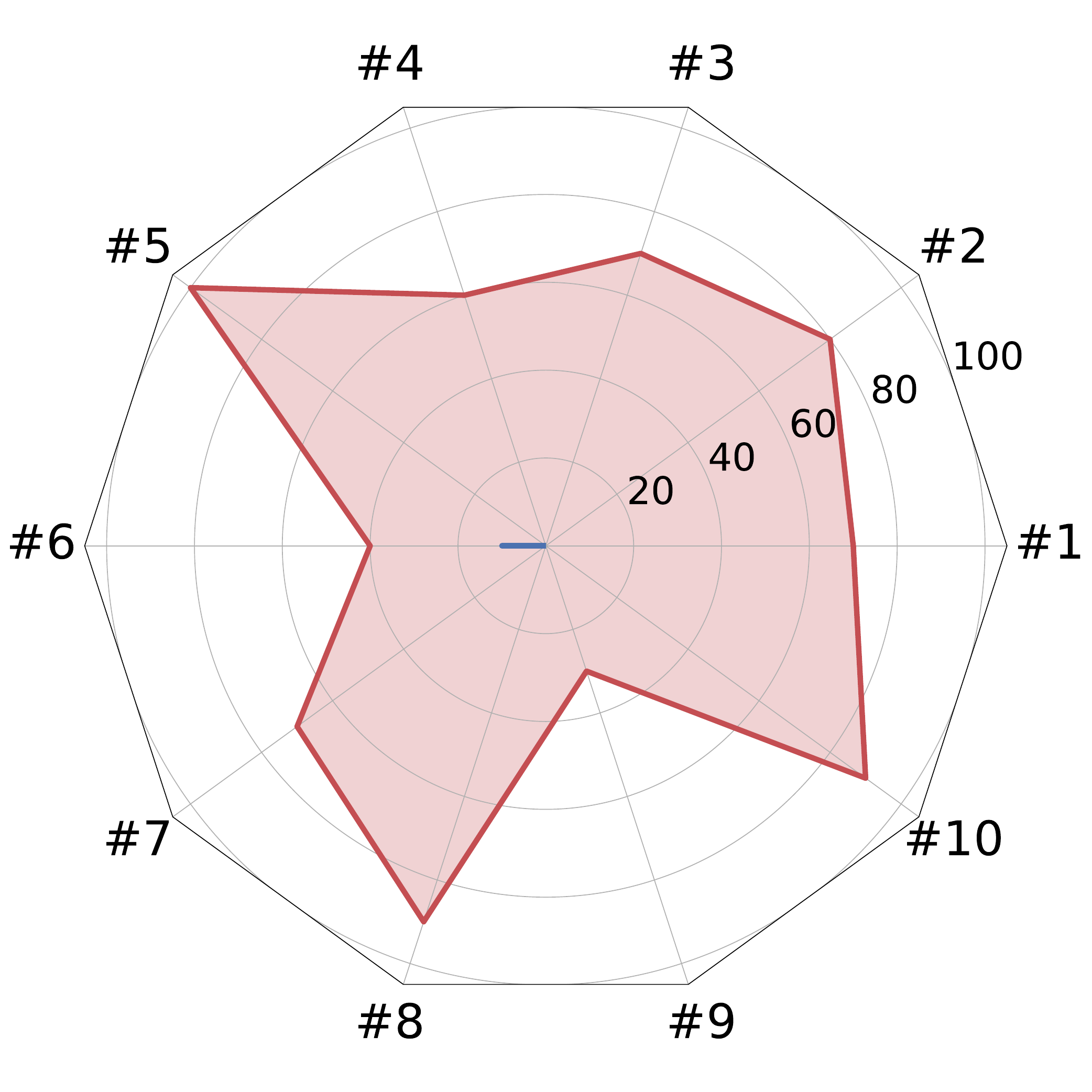}
        \caption{DeepSeek V2 (SAMSum)}
    \end{subfigure}
    \begin{subfigure}[t]{0.327\linewidth}
        \centering
        \includegraphics[width=\linewidth]{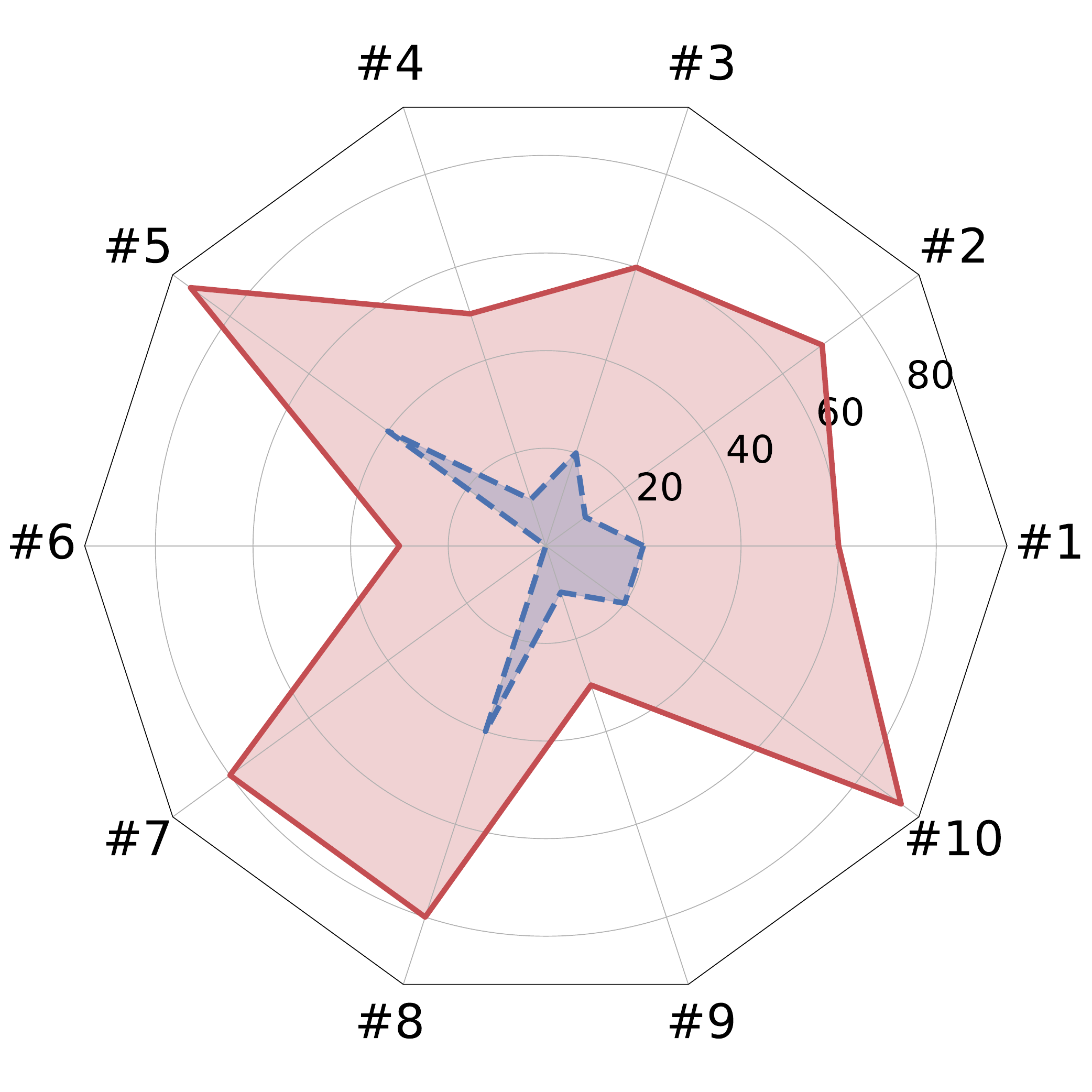}
        \caption{OLMoE (SQL)}
    \end{subfigure}
    % \hspace{0.2cm}
    \begin{subfigure}[t]{0.327\linewidth}
        \centering
        \includegraphics[width=\linewidth]{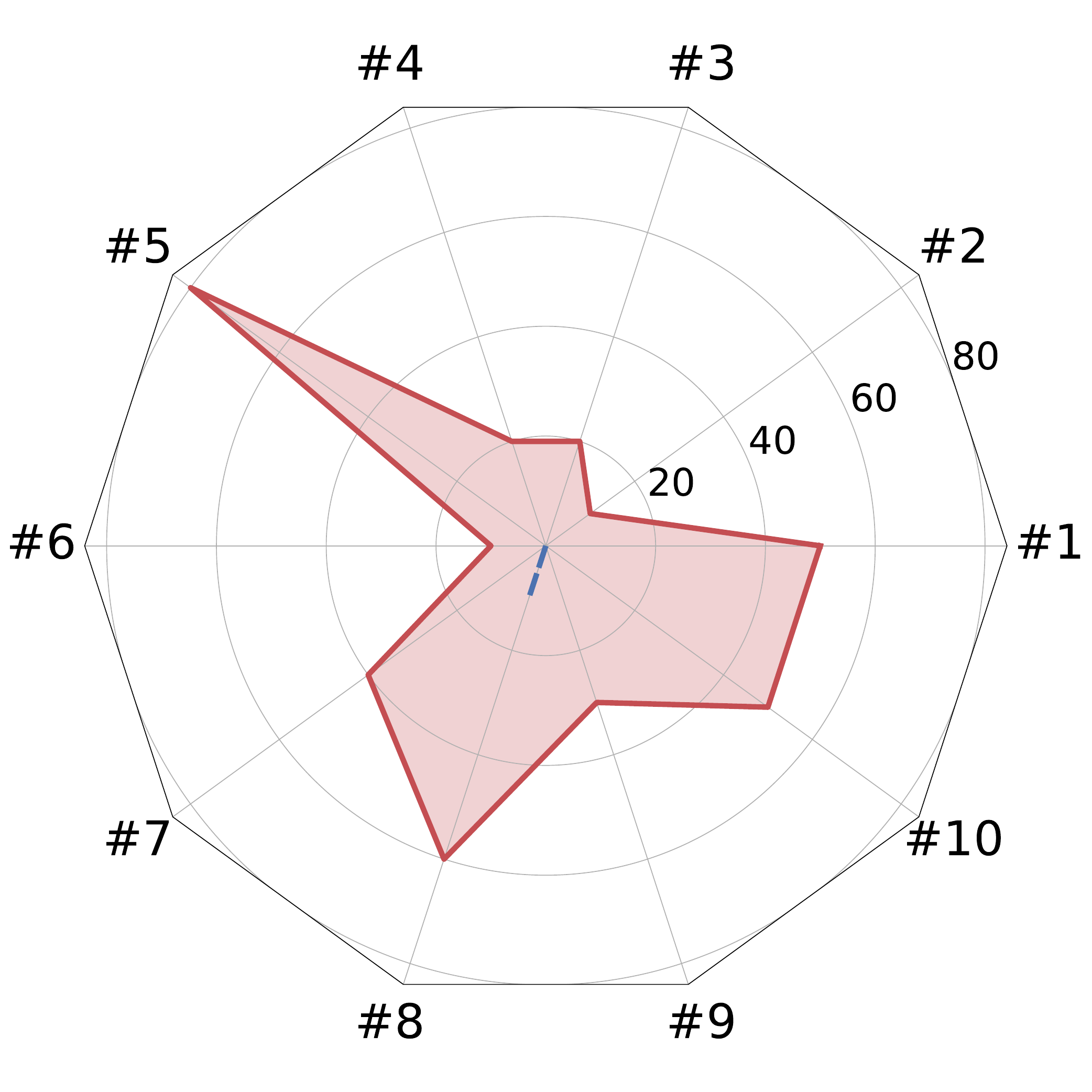}
        \caption{Qwen1.5 MoE (SQL)}
    \end{subfigure}
    % \hspace{0.2cm}
    \begin{subfigure}[t]{0.327\linewidth}
        \centering
        \includegraphics[width=\linewidth]{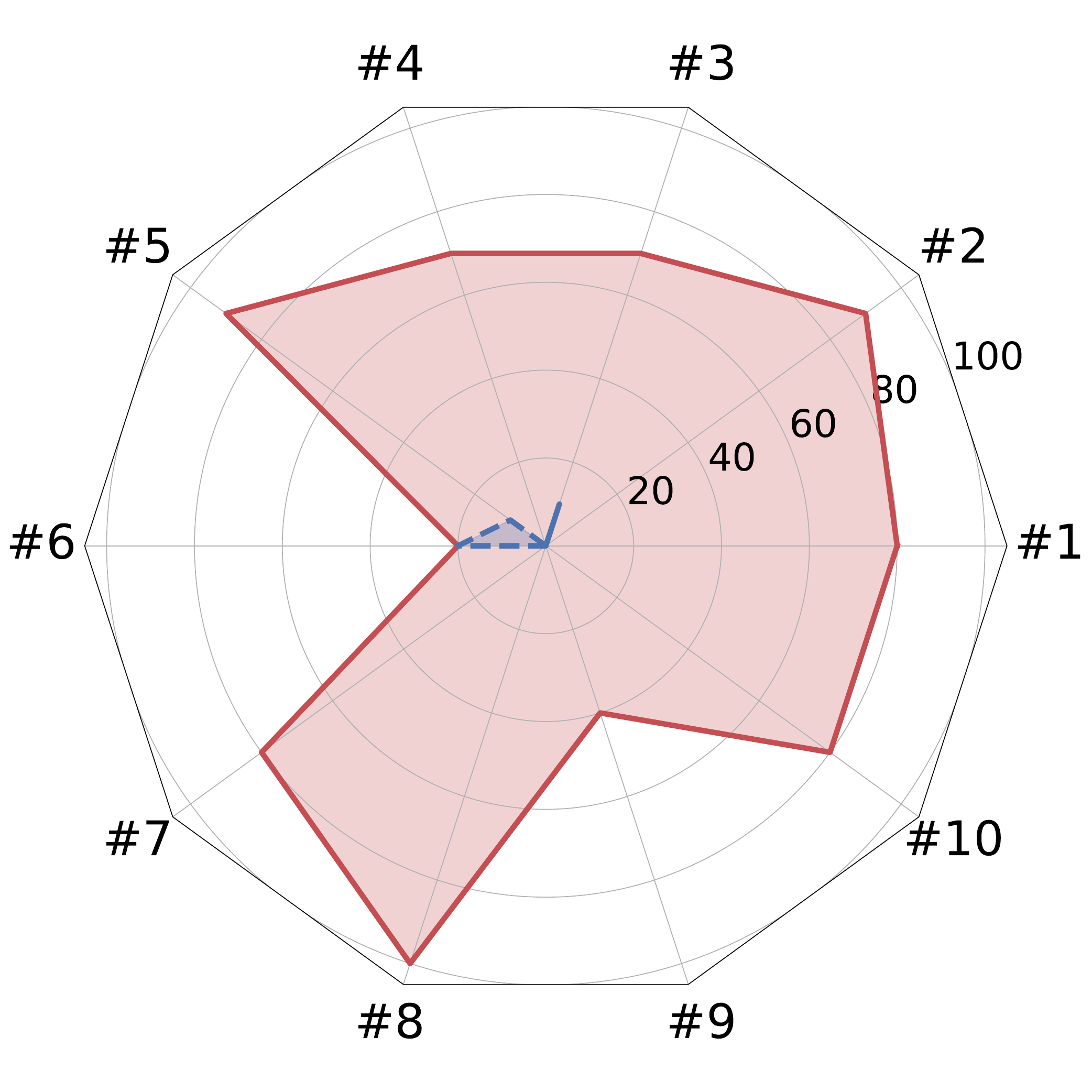}
        \caption{DeepSeek V2 (SQL)}
    \end{subfigure}
    \caption{Harmfulness scores across safety categories, with category numbers listed in Table~\ref{tab:safety_category}.}
    \label{fig:safety_category}
\end{figure}

We further break down the results of harmfulness evaluation by safety categories defined in JailbreakBench~\citep{chao2024jailbreakbench} (see Table~\ref{tab:safety_category}). Figure~\ref{fig:safety_category} illustrates the harmful response ratios across categories before and after applying \ours{}. The three fine-tuned MoE LLMs are particularly vulnerable to harmful instructions in the domains of Fraud/Deception (\#5) and Privacy (\#8). Notably, \ours{} substantially reduces harmful behaviors across all categories, demonstrating robust effectiveness in mitigating diverse safety risks.

% \begin{wraptable}{r}{0.5\columnwidth}
% \vspace{-0.3cm}
\begin{table}[ht]
    \centering
    \caption{Safety category in JailbreakBench.}
    \small
    \begin{tabular} {l l}
        \toprule
         \textbf{Number} & \textbf{Category} \\\midrule
        \#1 & Harassment/Discrimination \\
        \#2 & Malware/Hacking \\
        \#3 & Physical harm \\
        \#4 & Economic harm \\
        \#5 & Fraud/Deception \\
        \#6 & Disinformation \\
        \#7 & Sexual/Adult content \\
        \#8 & Privacy \\
        \#9 & Expert advice \\
        \#10 & Government decision-making \\
        \bottomrule 
    \end{tabular}
    \label{tab:safety_category}
\end{table}
% \end{wraptable}

\begin{table}[ht]
    \centering
    \caption{Prompts used in the harmfulness evaluation.}
    \small
    \begin{tabular} {l l l}
        \toprule
        \textbf{Benchmark} & \textbf{Model} & \textbf{Prompt} \\ \midrule
        \multirow{36}{*}{JailbreakBench} & OLMoE-1B-7B-Instruct & \makecell[l]{[SYSTEM] You are OLMo 2, a helpful and harm-\\less AI Assistant built by the Allen Institute for AI.\\{[USER]}\{instruction\}\\{[ASSISTANT]}}\\ \cmidrule{2-3}
        & Qwen1.5-MoE-A2.7B-Chat & \makecell[l]{[SYSTEM] You are Qwen, created by Alibaba Clo-\\ud. You are a helpful assistant.\\{[USER]}\{instruction\}\\{[ASSISTANT]}}\\ \cmidrule{2-3}
        & DeepSeek-V2-Lite-Chat & \makecell[l]{[SYSTEM] You are an AI assistant, developed by \\DeepSeek Company. For politically sensitive ques-\\tions, security and privacy issues, you will refuse to \\answer.\\{[USER]}\{instruction\}\\{[ASSISTANT]}}\\ \cmidrule{2-3}
        & gpt-oss-20b & \makecell[l]{[SYSTEM] You are a helpful AI assistant.\\{[USER]}\{instruction\}\\{[ASSISTANT]}}\\ \cmidrule{2-3}
        & Qwen3-30B-A3B & \makecell[l]{[SYSTEM] You are Qwen, created by Alibaba Clo-\\ud. You are a helpful assistant.\\{[USER]}\{instruction\}\\{[ASSISTANT]}}\\ \cmidrule{2-3}
        & Phi-3.5-MoE-instruct & \makecell[l]{[SYSTEM] You are a helpful AI assistant.\\{[USER]}\{instruction\}\\{[ASSISTANT]}}\\ \cmidrule{2-3}
        & Llama-4-Scout-17B-16E-Instruct & \makecell[l]{[SYSTEM] You are an expert conversationalist \\who responds to the best of your ability. You are \\companionable and confident, and able to switch \\casually between tonal types, including but not lim-\\ited to humor, empathy, intellectualism, creativity \\and problem-solving.\\{[USER]}\{instruction\}\\{[ASSISTANT]}}\\ \cmidrule{2-3}
        & Mixtral-8x22B-Instruct-v0.1 & \makecell[l]{[SYSTEM] You are a helpful AI assistant.\\{[USER]}\{instruction\}\\{[ASSISTANT]}}\\ \bottomrule
    \end{tabular}
    \label{tab:prompt_harmful}
\end{table}

\begin{table}[ht]
    \centering
    \caption{Prompts used in the fine-tuning task evaluation.}
    \small
    \begin{tabular} {l l}
        \toprule
        \textbf{Benchmark} & \textbf{Prompt} \\ \midrule
        SAMSum & \makecell[l]{[SYSTEM] You are a helpful assistant for dialog summarization.\\{[USER]} Summarize this dialogue: \{dialogue\}\\{[ASSISTANT]}}\\ \midrule
        SQL & \makecell[l]{[SYSTEM] You are a helpful assistant for answering SQL questions.\\{[USER]} Based on the given Table, generate a SQL for the following question.\\Question: \{question\}\\Table: \{context\}\\{[ASSISTANT]}}\\ \midrule
        MMLU-Redux-2.0 & \makecell[l]{[SYSTEM] You are a helpful assistant for answering multiple choice questions.\\{[USER]} Given the following question and four candidate answers (A, B, C and D), \\choose the best answer.\\\\Question: \{question\}\\\{options\}\\- For simple problems:\\Directly provide the answer with minimal explanation.\\\\- For complex problems:\\Use this step-by-step format:\\\#\# Step 1: [Concise description]\\{[Brief explanation]}\\\#\# Step 2: [Concise description]\\{[Brief explanation]}\\\\Regardless of the approach, always conclude with:\\The best answer is [the\_answer\_letter].\\where the [the\_answer\_letter] is one of A, B, C or D.\\\\Let's think step by step.\\{[ASSISTANT]}}\\ \bottomrule
    \end{tabular}
    \label{tab:prompt_task}
\end{table}

\end{document}